\documentclass{aastex62}

\usepackage{amsmath}	% Advanced maths commands
\usepackage{threeparttable} %need this so where put in table notes actually works

\newcommand{\lp}{\left(}
\newcommand{\rp}{\right)}
\newcommand{\lb}{\left[}
\newcommand{\rb}{\right]}

\newcommand{\msky}{m_{\rm{sky}}}

\newcommand{\Neff}{N_{\rm{eff}}}
\newcommand{\Neffhat}{\hat{N}_{\rm{eff}}}

\newcommand{\Ps}{P_{\rm{sc}}}
\newcommand{\texp}{t_{\rm{exp}}}

\newcommand{\seeing}{\sigma_{\rm{see}}}
\newcommand{\rhalf}{r_{\rm{half}}}

\newcommand{\ebv}{E(B-V)}

\newcommand{\texpo}{t_{\rm{exp, 0}}}

\newcommand{\Neffo}{N_{\rm{eff, 0}}}
\newcommand{\Kco}{K}
\newcommand{\Aco}{A}
\newcommand{\zpo}{ZP_0}
\newcommand{\zpt}{ZP}
\newcommand{\mskyo}{m_{\rm{sky, 0}}}

\newcommand{\logten}{\log_{\rm{10}}}
\newcommand{\Ne}{N_{\rm{e-}}}

\newcommand{\mAB}{m_{\rm{AB}}}

\newcommand{\PSmag}{m_{\rm{PS1}}}

\newcommand{\sigmasky}{\sigma_{\rm{sky}}}
\newcommand{\sigmaskyeff}{\sigma_{\rm{sky, eff}}}
\newcommand{\reltransp}{T_{\rm{rel}}}

\newcommand{\mdepth}{m_{\rm{depth}}}

\newcommand{\gb}{$g$}
\newcommand{\rband}{$r$}
\newcommand{\zb}{$z$}

\received{2020 February 13}
\revised{2020 May 12}
\accepted{2020 May 15}
\published{2020 July 9}
\submitjournal{AJ}

\shorttitle{Dynamic Observing and Tiling Strategies for the DESI Legacy Surveys}
\shortauthors{Burleigh et al.}
\begin{document}

\title{Dynamic Observing and Tiling Strategies for the DESI Legacy Surveys}

\correspondingauthor{Martin Landriau}
\email{mlandriau@lbl.gov}

\author{Kaylan J. Burleigh}
\affil{Department of Astronomy, University of California at Berkeley \\
501 Campbell Hall \#3411, Berkeley, CA 94720, USA}
\affil{Lawrence Berkeley National Laboratory \\
One Cyclotron Road, Berkeley, CA 94720, USA}

\author{Martin Landriau}
\affiliation{Lawrence Berkeley National Laboratory \\
One Cyclotron Road, Berkeley, CA 94720, USA}

\author{Arjun Dey}
\affiliation{NSF's Optical--Infrared Astronomy Research Laboratory\\
P.O. Box 26732, Tucson, AZ 85719, USA}

\author{Dustin Lang}
\affil{Perimeter Institute for Theoretical Physics, 31 Caroline Street N, Waterloo, Ontario, N2L 2Y5, Canada}
\affil{Department of Physics and Astronomy, University of Waterloo, Waterloo, ON N2L 3G1, Canada}

\author{David J. Schlegel}
\affiliation{Lawrence Berkeley National Laboratory \\
One Cyclotron Road, Berkeley, CA 94720, USA}

\author{Peter E. Nugent}
\affil{Department of Astronomy, University of California at Berkeley \\
501 Campbell Hall \#3411, Berkeley, CA 94720, USA}
\affil{Lawrence Berkeley National Laboratory \\
One Cyclotron Road, Berkeley, CA 94720, USA}

\author{Robert Blum}
\affiliation{NSF's Optical--Infrared Astronomy Research Laboratory\\
P.O. Box 26732, Tucson, AZ 85719, USA}

\author{Joseph R. Findlay}
\affiliation{Department of Physics \& Astronomy\\
University of Wyoming, 1000 E. University, Dept. 3905, Laramie, WY 82071, USA}

\author{Douglas P. Finkbeiner}
\affiliation{Harvard-Smithsonian Center for Astrophysics\\
Harvard University, 60 Garden Street, Cambridge, MA 02138, USA}

\author{David Herrera}
\affiliation{NSF's Optical--Infrared Astronomy Research Laboratory\\
P.O. Box 26732, Tucson, AZ 85719, USA}

\author{Klaus Honscheid}
\affiliation{Department of Physics, Ohio State University\\
191 West Woodruff Avenue, Columbus, OH 43210, USA}

\author{St\'{e}phanie Juneau}
\affiliation{NSF's Optical--Infrared Astronomy Research Laboratory\\
P.O. Box 26732, Tucson, AZ 85719, USA}

\author{Ian McGreer}
\affiliation{Steward Observatory, University of Arizona\\
933 N. Cherry Avenue, Tucson, AZ 85721, USA}

\author{Aaron M. Meisner}
\affiliation{NSF's Optical--Infrared Astronomy Research Laboratory\\
P.O. Box 26732, Tucson, AZ 85719, USA}

\author{John Moustakas}
\affil{Department of Physics \& Astronomy, Siena College\\
515 Loudon Road, Loudonville, NY 12211, USA}

\author{Adam D. Myers}
\affiliation{Department of Physics \& Astronomy\\
University of Wyoming, 1000 E. University, Dept. 3905, Laramie, WY 82071, USA}

\author{Anna Patej}
\affiliation{Stanford Law School, 559 Nathan Abbott Way, Stanford, CA 94305, USA}

\author{Edward F. Schlafly}
\affil{Lawrence Livermore National Laboratory \\
7000 East Avenue, Livermore, CA 94550, USA}

\author{Francisco Valdes}
\affiliation{NSF's Optical--Infrared Astronomy Research Laboratory\\
P.O. Box 26732, Tucson, AZ 85719, USA}

\author{Alistair R. Walker}
\affiliation{Cerro Tololo Inter-American Observatory, NSF's Optical--Infrared Astronomy Research Laboratory\\
Casilla 603, La Serena, Chile}

\author{Benjamin A. Weaver}
\affiliation{NSF's Optical--Infrared Astronomy Research Laboratory\\
P.O. Box 26732, Tucson, AZ 85719, USA}

\author{Christophe Y\`{e}che}
\affiliation{CEA, Centre de Saclay, IRFU/DPhP,  F-91191 Gif-sur-Yvette, France}

\author{the DECaLS, MzLS, and BASS Teams}



\begin{abstract}

The Dark Energy Spectroscopic Instrument Legacy Surveys, a combination of three
ground-based imaging surveys, have mapped 16,000 deg$^2$ in three
optical bands ($g$, $r$, and $z$) to a depth 1--$2$~mag deeper than
the Sloan Digital Sky Survey.  Our work addresses one of the
major challenges of wide-field imaging surveys conducted at
ground-based observatories: the varying depth that results from
varying observing conditions at Earth-bound sites.  To mitigate these
effects, the Legacy Surveys (the Dark Energy
Camera Legacy Survey, or DECaLS; the Mayall
$z$-band Legacy Survey, or MzLS; and the
  Beiijing-Arizona Sky Survey, or BASS) employed a unique strategy to
dynamically adjust the exposure times as rapidly as possible in
response to the changing observing conditions. We present the tiling
and observing strategies used by the first two of these surveys. We demonstrate that
the tiling and dynamic observing strategies jointly result in a more
uniform-depth survey that has higher efficiency for a given total
observing time compared with the traditional approach of using fixed
exposure times.

\end{abstract}



\section{Introduction}
\label{sec:intro}

The Legacy Surveys\footnote{\url{http://legacysurvey.org}}
\citep[see][]{bassOverview, Paper1}
 are a combination of three imaging surveys that have mapped two
contiguous areas totaling $16{,}000$~deg$^2$ in three optical bands
($g$, $r$ and $z$) to depths 1--2~mag deeper than the Sloan Digital
Sky Survey imaging \citep[SDSS; e.g.][]{sdssDR7}. The three surveys
that make up the Legacy Surveys are the DECam Legacy Survey (DECaLS),
the Mayall \zb-band Legacy Survey (MzLS), and the Beijing-Arizona Sky
Survey (BASS). DECaLS uses the Blanco 4m telescope and Dark Energy
Camera
\citep[DECam;][]{decam}
located at Cerro Tololo, Chile; MzLS uses the Mosaic3 camera
\citep{mosaic3} at the Mayall Telescope located at Kitt Peak in
Arizona; and
BASS uses the Bok 2.3m telescope/90Prime camera on Kitt
Peak
\citep{90Prime}. MzLS was completed in early 2018 and the other two
surveys were completed in early 2019.  The Legacy Surveys also provide
mid-infrared photometric measurements for all optical sources derived
using forced photometry on coadded multi-epoch data from the
Wide-field Infrared Survey Explorer \citep[WISE;][]{wright10}.

The primary purpose of the Legacy Surveys is to provide targets for
the Dark Energy Spectroscopic Instrument
\citep[DESI;][]{desiScience,desiInstrument}. DESI is a robotically
actuated 5000-fiber spectrograph that will survey 14,000 deg$^2$ of
sky in order to  make a Stage IV measurement of dark energy. Spectra
and redshifts of more than 30 million galaxies and quasars will be
obtained over this five-year survey.
DESI, installed at the prime focus of the Mayall 4m telescope on Kitt
Peak, Arizona, is currently being commissioned and will begin survey
operations in 2021.

In addition to providing targets for DESI, the Legacy Surveys have already
dramatically improved the utility of existing spectroscopic and imaging
datasets, by spanning the SDSS footprint and being  1--2 mag deeper,
with better image quality, than either the SDSS or the Panoramic
Survey Telescope and Rapid Response System 1 (Pan-STARRS 1, or PS1)
$3\pi$ survey \citep{panstarrs}. Existing spectroscopic datasets in
the DESI footprint include SDSS, the Two-degree-Field Galaxy Redshift
Survey, and the WiggleZ Dark Energy Survey.
Increasing $g-$, $r-$, and $z$-band depths
by 1.5--2 mag, increases the number of $z > 0.5$ galaxies with
imaging measurements by about a factor of 30. No other ongoing or
currently planned ground-based survey will provide this depth or
uniformity of depth over as large an extragalactic ({\it i.e.} $\vert b
\vert \ge 20^\circ$) footprint as the Legacy Surveys, especially in
the northern hemisphere. For example, the Dark Energy Survey has
observed
$5000$~deg$^2$ of the southern sky, overlapping only about 1000
deg$^2$ of the SDSS footprint \citep{des}.

All previous ground-based wide-field imaging surveys have used fixed
exposure times per band, which result in survey depths that vary
across the survey footprint due to both terrestrial and
extraterrestrial constraints. Terrestrial constraints include the
observing conditions (i.e., cloud cover, transparency, delivered image
quality, sky brightness) and telescope limitations (e.g., zenith
distance of observation, telescope pointing accuracy, telescope
tracking accuracy, focus, etc.). Extraterrestrial constraints include
the extinction due to Galactic and solar system dust, zodiacal dust,
sky brightness, Galactic cirrus and other sources of diffuse emission,
and source crowding. Cosmological surveys require a uniformity of
depth over a large area, and hence imaging surveys with varying depth
are generally truncated near their shallowest depth or are subjected
to uncertain completeness corrections.

In this paper, we describe the innovative approach employed in
our observing strategy for DECaLS and MzLS (the observing strategy for
BASS is presented in \citealt{bassOverview}). Instead of adopting a
fixed exposure time, we analyzed images contemporaneously in order to
dynamically adjust the exposure time to ensure a near-constant depth
for each image. This procedure allowed us to optimally use the
available telescope time with the minimum of reobservation. This
optimization was particularly important given that the
imaging surveys had to be completed to a minimum depth in less than
four years due to the DESI construction and installation schedule.

The paper is organized as follows. In section~\ref{tiling}, we present
the choices of tiling for DECaLS and MzLS. In
section~\ref{sec:nightlyplan}, we describe the goals of our observing
strategy. In section~\ref{dynamic}, we discuss how we implemented
dynamic observing.

\section{Tiling Strategy \label{tiling}}

\subsection{General Concepts}

Wide-field imaging surveys typically aim to cover one or more
contiguous areas of sky much larger than the footprint of the imaging
camera. The Legacy Surveys represent a particularly extreme case where
we have imaged a 16,000~deg$^2$ region using cameras that
have fields of view of between $0.36$ and $3.18$~deg$^2$ (see
Table~\ref{tab:cameras}). In addition, all of the camera focal planes
are CCD mosaics that have gaps between individual CCDs. Hence, an
efficient tiling pattern has to both cover the entire area with as few
tiles as possible, and also cover all of the CCD gaps to some minimum
depth driven by the science requirements. 

Once the basic tiling strategy was identified, we defined a total of
three independent tilings, with each tiling offset from the other two
by some prescribed amount. Three tilings ensure that the footprint is
almost entirely covered, while also minimizing the amount of area
that does not have at least two images at any given position. Two-pass
coverage is useful both to discriminate and mask any particle events
or other detector-based anomalies, and to boost the signal-to-noise
ratio $(S/N)$
compared to a single pass. We used a Monte Carlo process of different
offsets for the tiling sets for each camera in order to select the
offsets that maximized three-pass coverage while minimizing one-pass
coverage.

The detailed implementations for each camera are described in the
following two subsections.

\begin{table*}
\begin{threeparttable}[b]
 \caption{Camera Properties}
 \label{tab:cameras}
 \begin{tabular}{lccccccc}
  \hline
  Camera & CCDs & Amplifiers & Pixels & Pixel & FOV & Fill\\
  &  & (per CCD) & (per CCD) & Scale & (deg$^2$) & Factor\\
  \hline
  DECam & 62 & 2 & 4094 x 2046 & 0.262 & 3.18 %($\sim$round) 
  & 0.87 \\
  Mosaic3 & 4 & 4 & 4079 x 4054 & 0.260 & 0.36 & 0.95\\
  90Prime & 4 & 4 & 4096 x 4032 & 0.455 & 1.16$\times$1.16 & 0.94 \\
%  WISE 0.4-m & $--$ & W1, W2 & all sky & all sky & $--$ & Jan. 2010 & 2017 & 90\%  \\
  \hline
 \end{tabular}
 Pixel scale: arcsec pixel$^{-1}$. 
 FOV: camera field of view including CCD gaps and dead CCDs.
 Fill factor: fraction of the field of view that is covered by CCDs.
 \begin{tablenotes}
 \item []
 \end{tablenotes}
  \end{threeparttable}
  \end{table*}
  
  %%%%%%%
\subsection{Implementation for DECaLS}

DECam has a roughly circular field of view of $a_{\rm FoV}$=3.18~deg$^2$
\citep{decam}.  To cover the entire sky, the ideal tiling would
require $N=4\pi(180/\pi)^2/a_{\rm FoV}=12973$ tiles (not accounting
for CCD gaps or nonworking CCDs, see
Table~\ref{tab:cameras}).

For defining the tiling for DECaLS, we adopted the approach of \citep{hardin2012}, who considered the
general problem of covering a sphere uniformly with a fixed number of
points.  We selected
the precomputed icosahedral arrangements
of Hardin et al. (2012) with tiling $N_{\rm tile}$ that was close in number
to but greater than $N$ (i.e., the minimum number while still
providing sufficient overlap with the neighboring tile).  We investigated the icosahedral tilings with
$N_{\rm tile}=[15,252,~15,392,~15,872,~16,002,~16,472,~16,752]$ and settled on
$N=15,872$ as providing the
best solution with 99.98\% and 98.01\% of the sky
having at least two and three exposures respectively, using a 3-pass strategy.

\begin{figure}
  \begin{center}
    \begin{tabular}{ccc}
      One pass & Two passes & Three passes \\
      \includegraphics[width=0.2\textwidth]{tile-00} &
      \includegraphics[width=0.2\textwidth]{tile-31} &
      \includegraphics[width=0.2\textwidth]{tile-62}
      \\
      \includegraphics[width=0.2\textwidth]{tile-06} &
      \includegraphics[width=0.2\textwidth]{tile-37} &
      \includegraphics[width=0.2\textwidth]{tile-68}
      \\
      \includegraphics[width=0.3\textwidth]{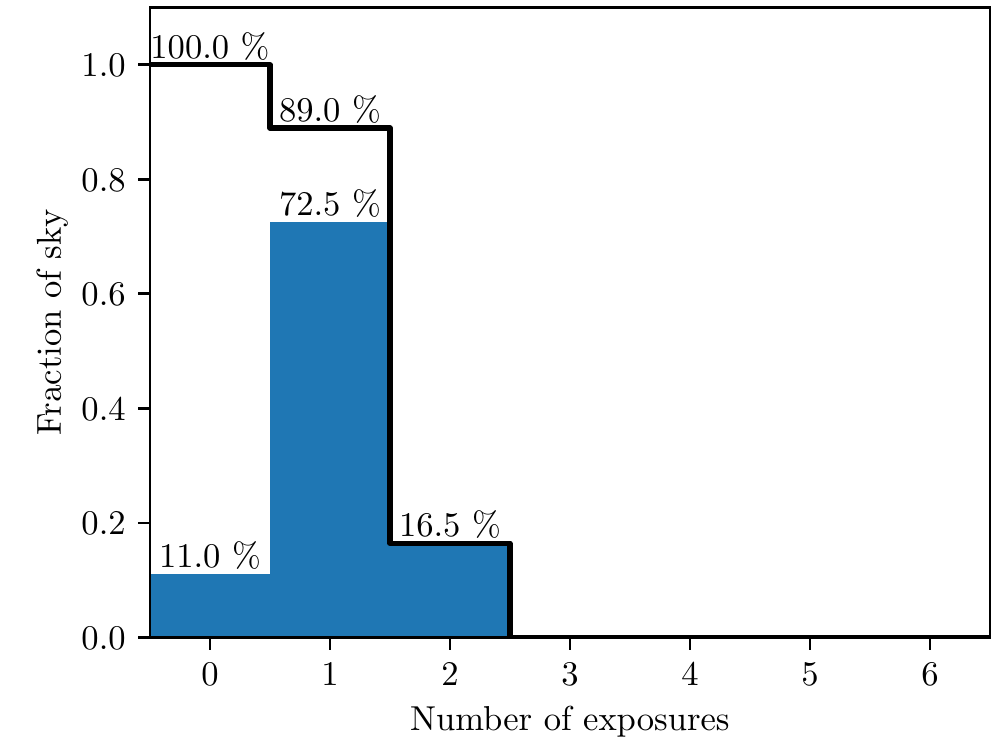} &
      \includegraphics[width=0.3\textwidth]{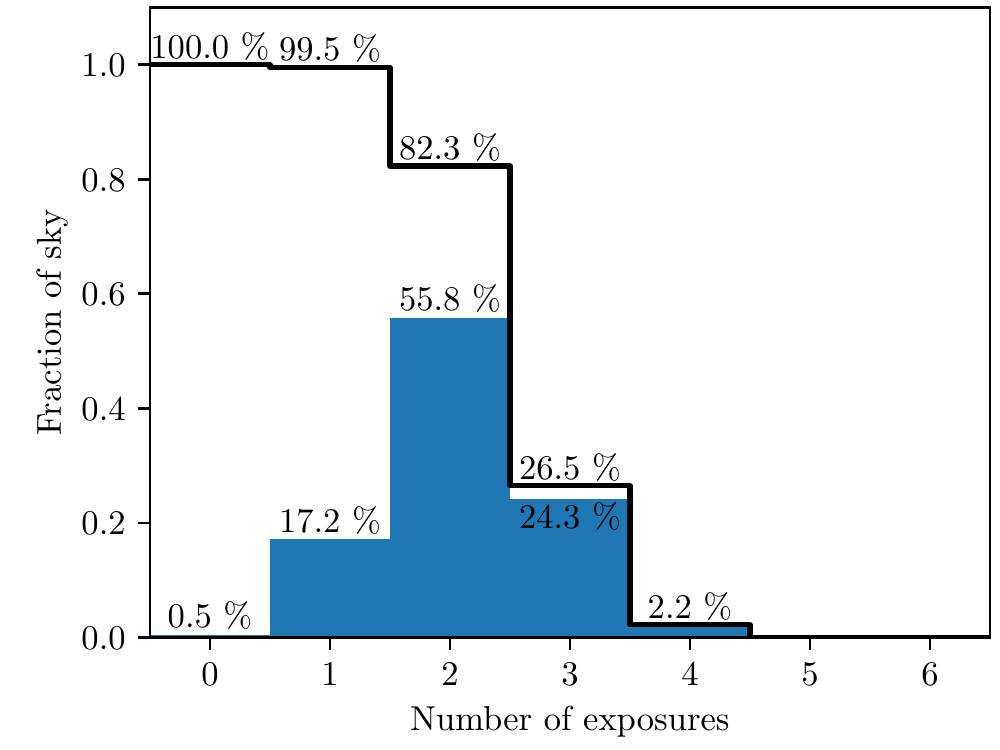} &
      \includegraphics[width=0.3\textwidth]{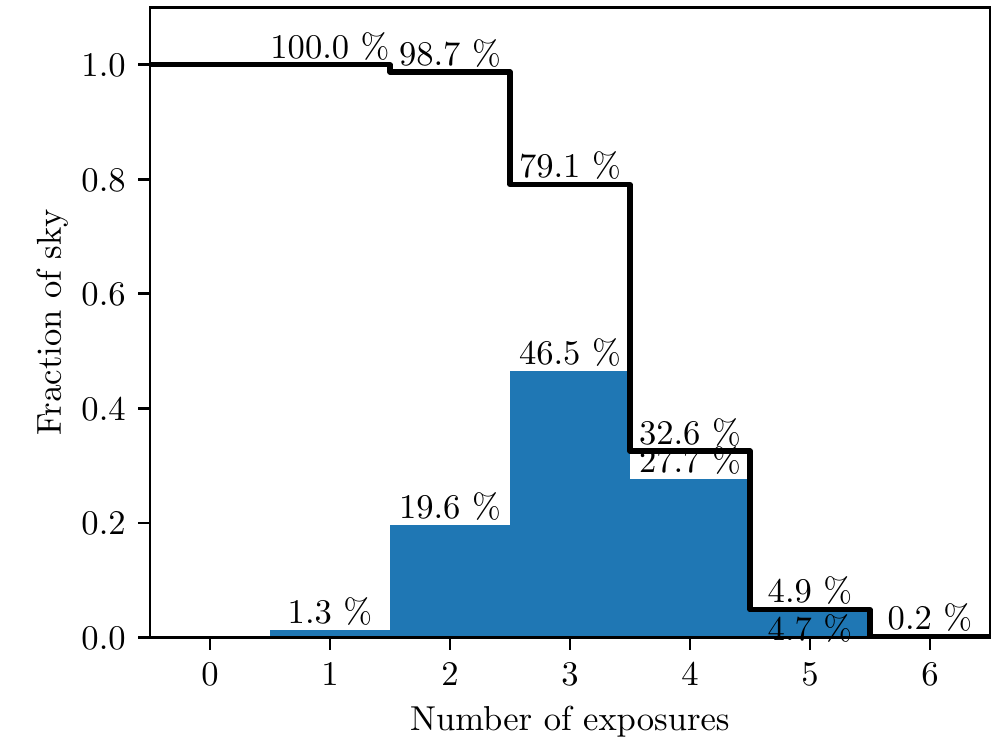}
      \\
    \end{tabular}
  \end{center}
  \caption{Tiling strategy in the DECaLS survey.  DECam has 62 science
    CCDs, but during the course of the survey, one or two CCDs have
    been inoperative.  In the example exposure shown, CCD N30 is
    inoperative, leaving a hole in the edge of the hexagonal
    footprint.  The first column shows a region of sky (about
    $5.5^{\circ}$ wide) covered with our Pass 1 tiling, with a
    single exposure in the top row and neighboring tiles in the second
    row.  The bottom row shows the approximate coverage statistics,
    where the x-axis represents the number of repeat exposures.
    The black lines should be compared to the numbers in Table
    \ref{tab:tiling} and show the fraction of sky that have at least
    $N+1$ exposures; the difference between the pass 3 numbers and
    those in the table result mainly from small pointing errors.  The
    blue histograms show the fraction of sky that only have $N+1$ exposures.
    The second and third columns show the coverage after our Pass 2
    and Pass 3 tilings have been added, respectively.
    \label{fig:tilingdecals}}
\end{figure}

Two of the three passes are copies of the first pass, offset by
[$\Delta$R.A., $\Delta$decl.] of [0.2917,~0.0833]\,deg
and [0.5861,~0.1333]\,deg respectively. This solution results in
fractional coverage within the DESI footprint as shown in
Table~\ref{tab:tiling}. Ideally, we would obtain three-image coverage
of 100\%, but this is not possible with a three-pass strategy given
the gaps between the DECam CCDs. The resulting tiling for DECaLS is
shown in Figure~\ref{fig:tilingdecals} along with the as-observed
coverage statistics (which include pointing errors during the
observations).

\subsection{Implementation for MzLS}
   
The Mosaic3 Camera has an
approximately square on-sky footprint with a field of view of
$35.89^\prime\times36.06^\prime$ \citep[Table~\ref{tab:cameras}; see
also ][]{mosaic3}. Given the smaller size and roughly square
footprint, we settled on a tiling pattern that was aligned along rows
of constant declination, with adjacent frames overlapping by $1.7^\prime$
on all four sides. The resulting map has 122{,}765 tile centers in a single
pass; the two other passes are offset by $11.7^\prime$ and $23.5^\prime$ in decl.,
respectively (or one third and two third of the field of view).

\begin{figure}
  \begin{center}
    %
    % THESE FIGURES COME FROM legacypipe: legacyanalysis/tiling.py : mosaic()
    %
    \begin{tabular}{ccc}
      One pass & Two passes & Three passes \\
      \includegraphics[width=0.2\textwidth]{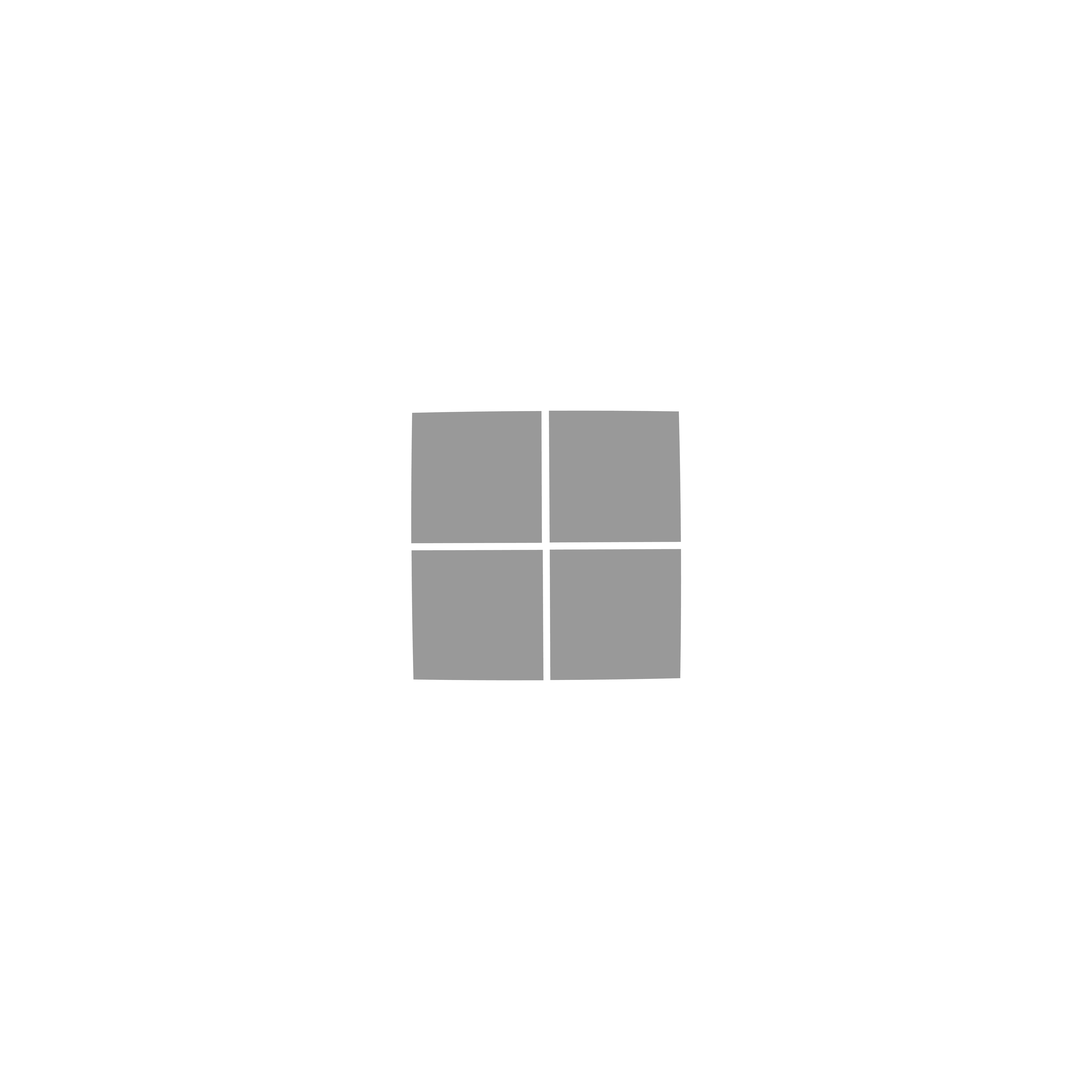} &
      \includegraphics[width=0.2\textwidth]{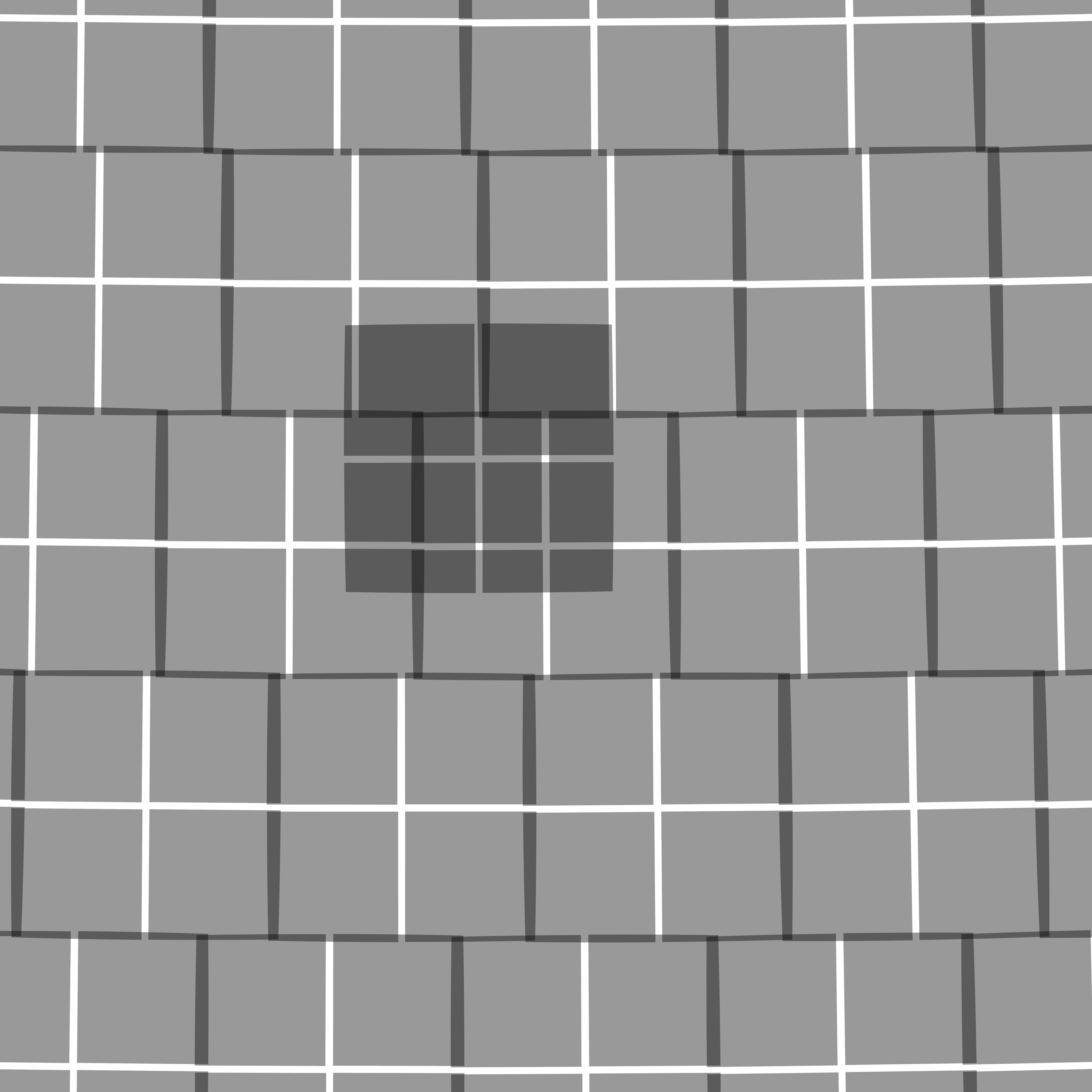} &
      \includegraphics[width=0.2\textwidth]{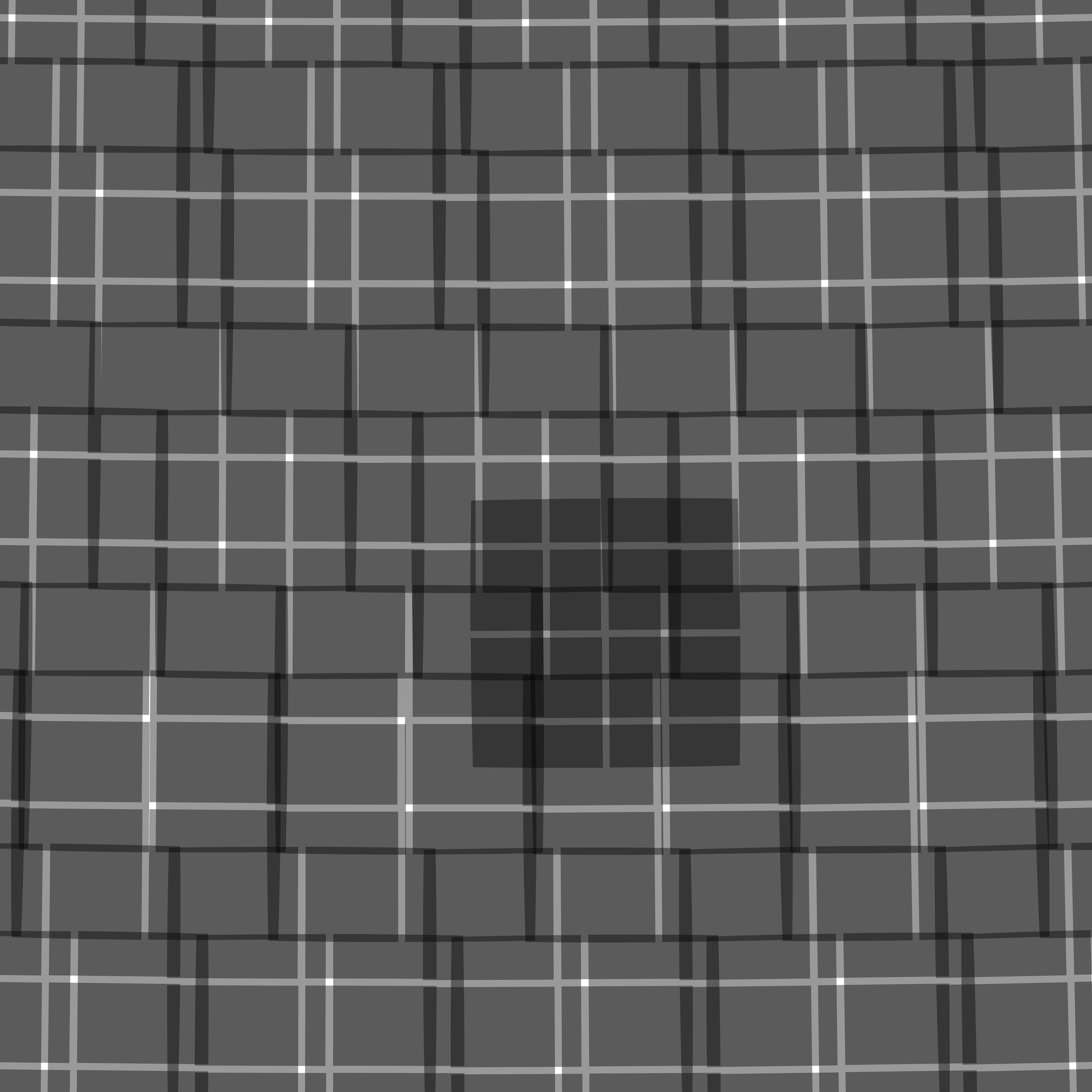}
      \\
      \includegraphics[width=0.2\textwidth]{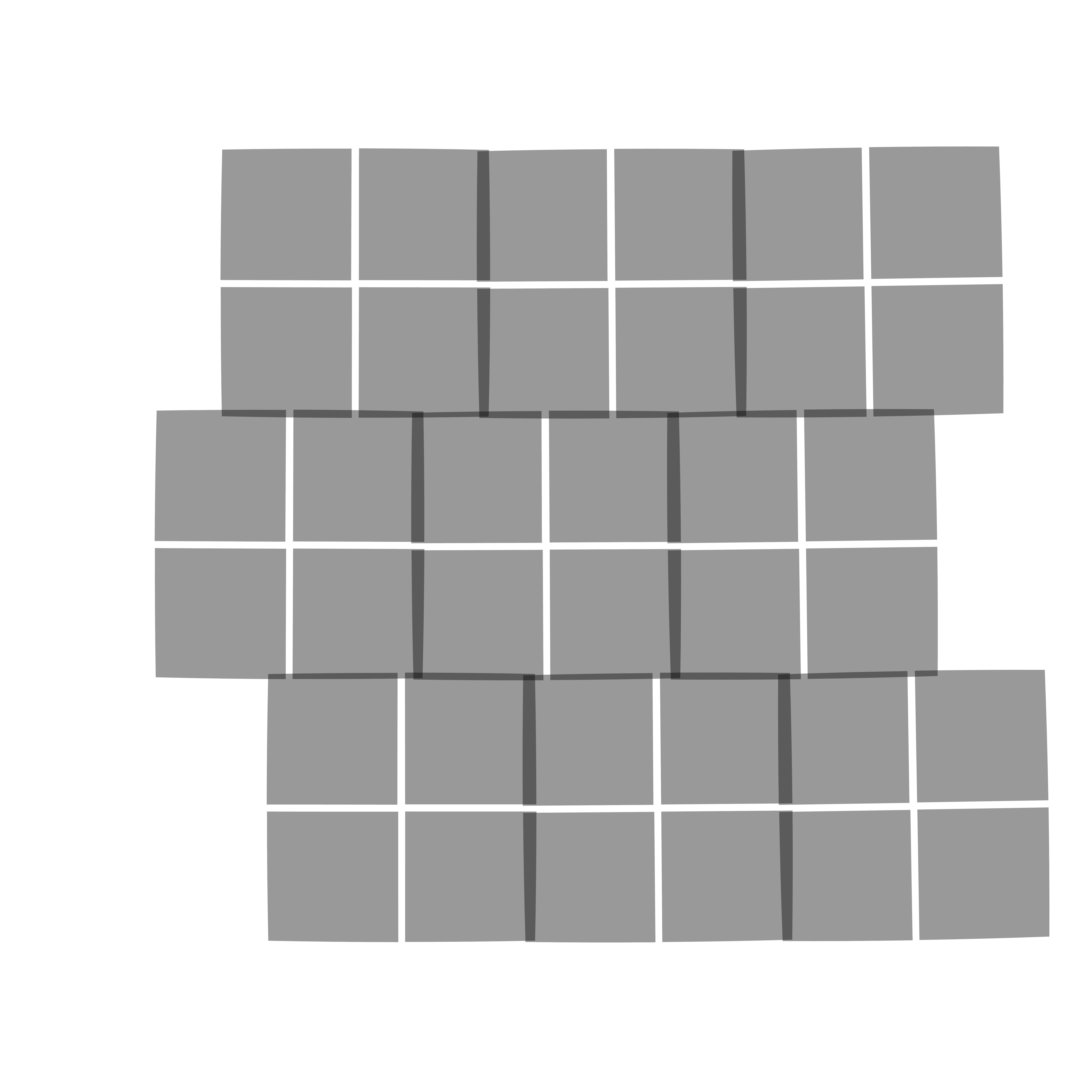} &
      \includegraphics[width=0.2\textwidth]{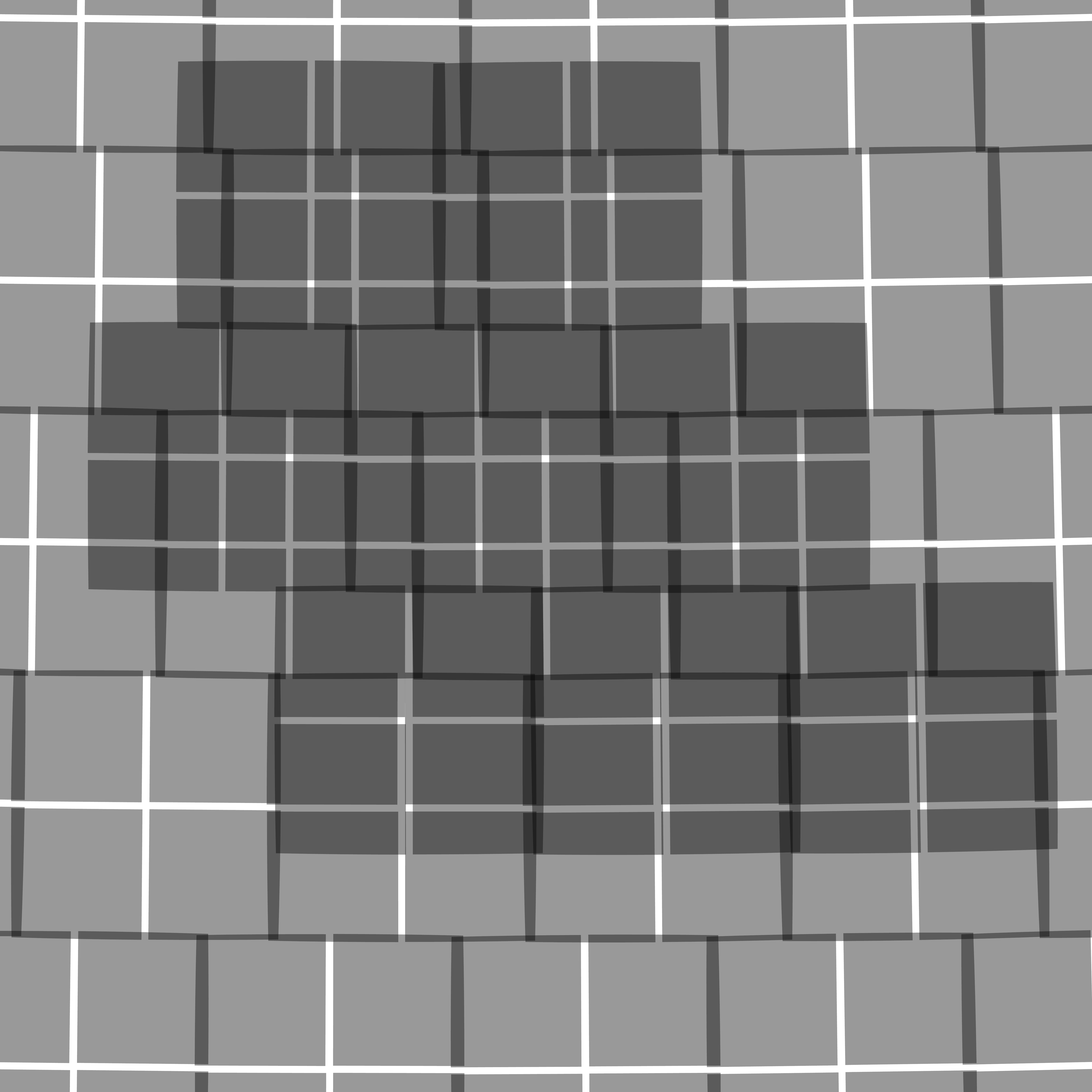} &
      \includegraphics[width=0.2\textwidth]{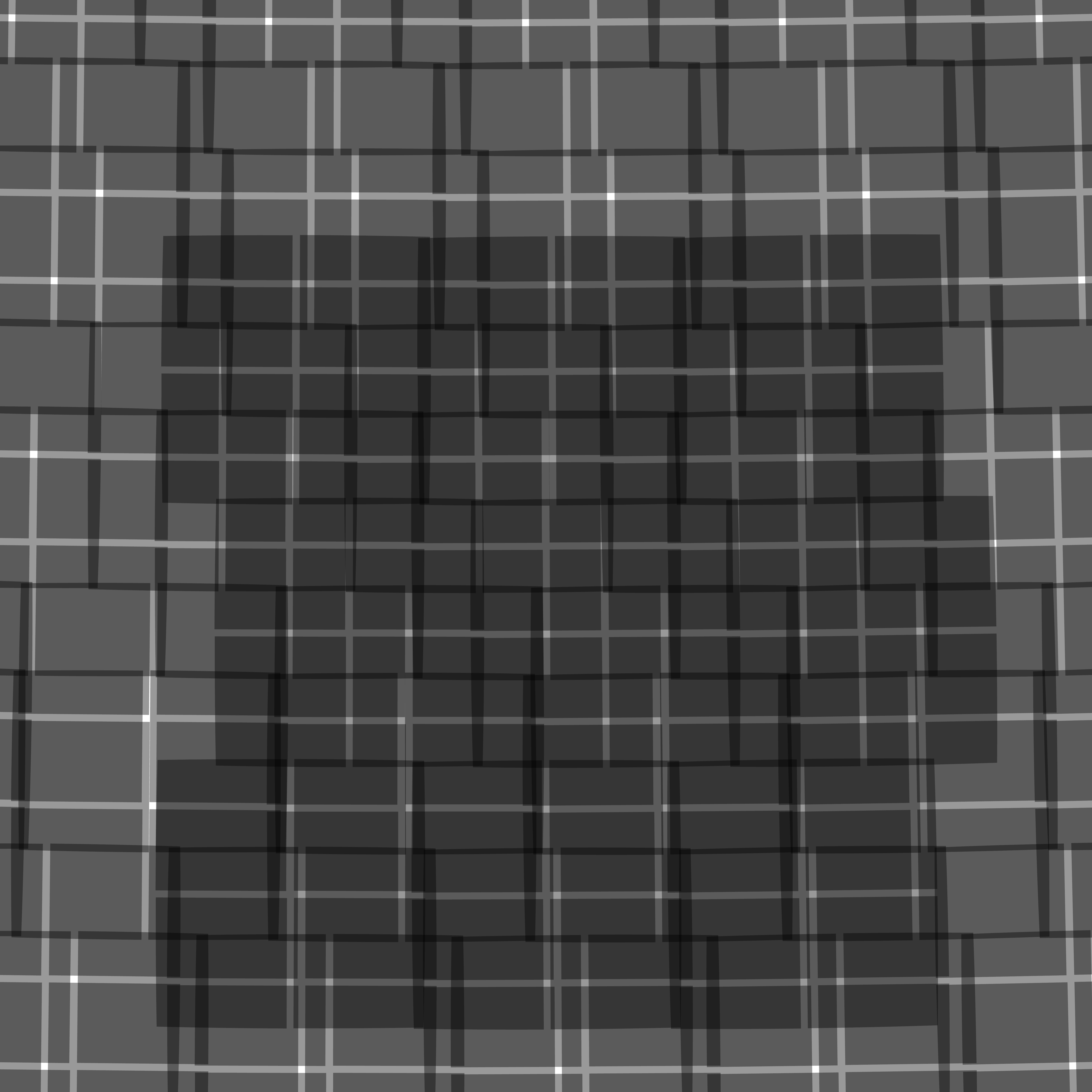}
      \\
      \includegraphics[width=0.3\textwidth]{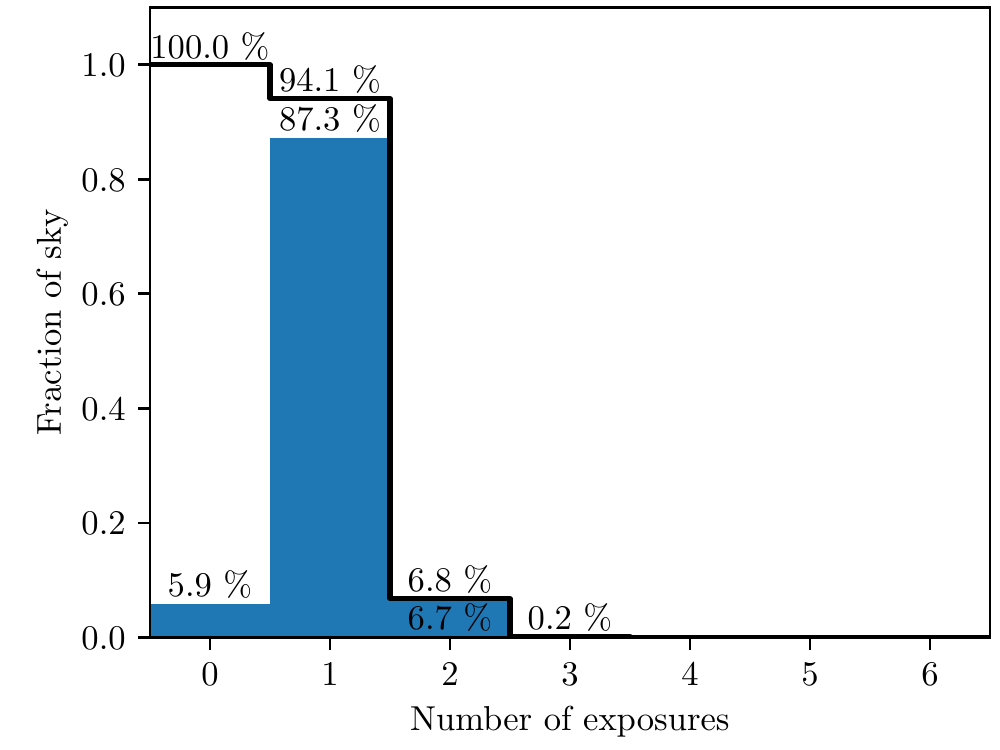} &
      \includegraphics[width=0.3\textwidth]{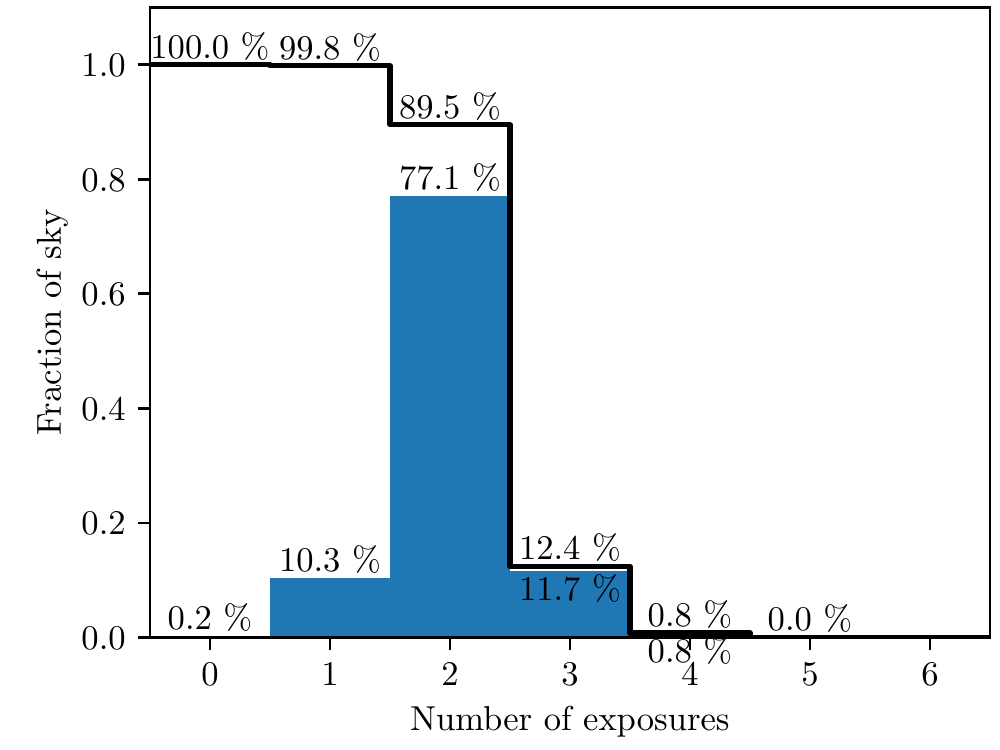} &
      \includegraphics[width=0.3\textwidth]{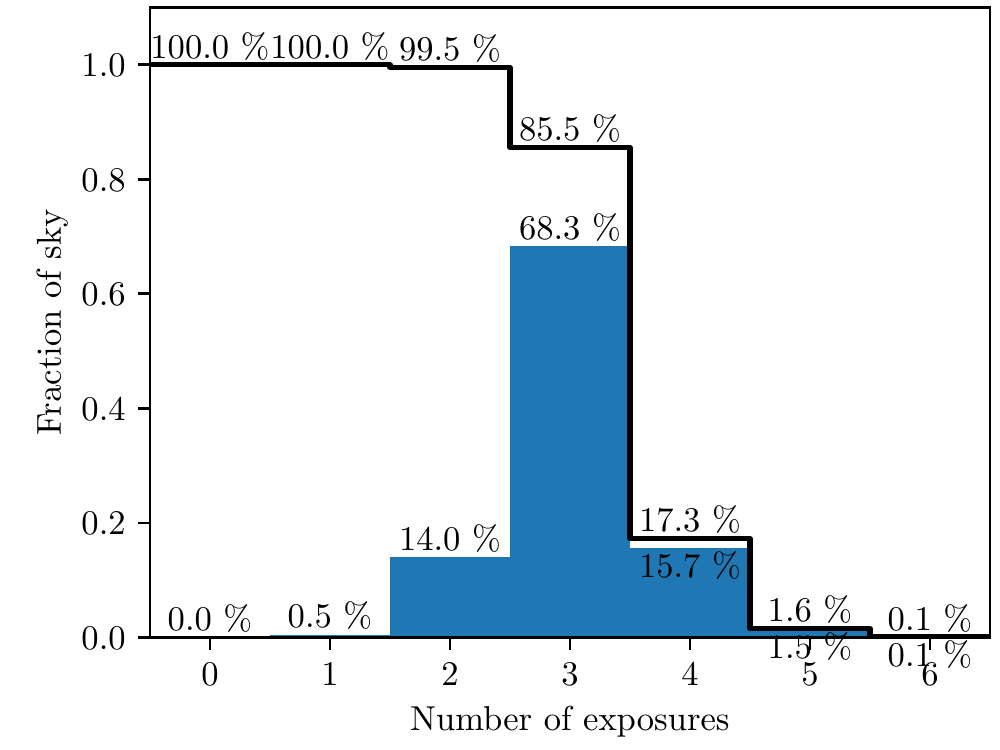}
      \\
    \end{tabular}
  \end{center}
  \caption{Tiling strategy in the MzLS survey.  The Mosaic3 camera has 4 CCDs,
    each with a field of view about $0.3^{\circ} \times 0.3^{\circ}$, with small gaps
    between the CCDs.
    The first column shows a region of sky (about
    $2.5^{\circ}$ wide) covered with our Pass 1 tiling, with a
    single exposure in the top row and neighboring tiles in the second
    row.  The bottom row shows the approximate coverage statistics,
    where the x-axis represents the number of repeat exposures.
    The black lines should be compared to the numbers in Table
    \ref{tab:tiling} and show the fraction of sky that have at least
    $N+1$ exposures; the difference between the Pass 3 numbers and
    those in the table result mainly from small pointing errors.  The
    blue histograms show the fraction of sky that only have $N+1$ exposures.
    The second and third columns show the coverage after our Pass 2
    and Pass 3 tilings have been added, respectively.
    \label{fig:tilingmzls}}
\end{figure}

This choice of tiling ensures that 99.5\% of the footprint is covered by at
least three exposures (see Table~\ref{tab:tiling}).  The tiling for MzLS is
shown in Figure~\ref{fig:tilingmzls} along with the as-observed
coverage statistics.

\begin{table}
\begin{center}
\begin{threeparttable}[b]
 \caption{Tiling Solutions for DECaLS and MzLS}
 \label{tab:tiling}
 \begin{tabular}{lccc}
  \hline
  $N$ & DECaLS & MzLS \\
  % & (c15872) & () \\
  \hline
  0 & 1.0000 & 1.0000 \\
  1 & 0.9998  & 1.0000 \\
  2 & 0.9801 & 0.9950 \\
  3 & 0.7443 & 0.8500 \\
  \hline
 \end{tabular}
Note. The DECaLS and MzLS columns are the fraction of the sky
footprint having a given number of repeat exposures ($N$).
%N: \\
%DECALS:  
 \begin{tablenotes}
 \item []
 \end{tablenotes}
  \end{threeparttable}
  \end{center}
  \end{table}

%%%%%%%
\section{Observing Strategy}
\label{sec:nightlyplan}

\subsection{Optimizing for Photometric Calibration and Image Quality}

Three passes, each constituting a complete tiling of the footprint as described in the previous section,
were chosen to maximize the scientific uniformity and utility of the
survey. In order to ensure that a given survey could be
photometrically calibrated, we reserved the first tiling of the
footprint (Pass 1) for times with photometric conditions when the
seeing was good (i.e., $<$1.3\arcsec). We reserved the second tiling
(Pass2) for times with {\it either} photometric conditions {\it
  or} good seeing. We reserved the third tiling (Pass 3) for times
when neither of these conditions were met, but were still deemed
acceptable, \textit{i.e.} seeing$<$2\arcsec,
  transparency $>$0.9 and sky brightness no more than 0.25 mag brighter
  than the nominal value for that band.
This strategy was designed to ensure that every point
within the survey footprint had at least one image that could be
photometrically calibrated and at least one image that had good
seeing.  We observed three passes across the entire survey footprint
to ensure the high completeness in two passes across every point in the
footprint.  During times when the weather was poor (i.e.,
worse than Pass 3 conditions), we still took data when possible,
but these data did not contribute to the final catalogs. 

\subsection{Optimizing the Nightly Plan}

As much as possible, we scheduled \zb-band observations during bright time
(i.e.\ when the Moon was above the horizon, or the Sun's altitude was
between $-10^{\circ}$ and $-15^{\circ}$) and reserved dark time
for \gb \, and \rband. With these constraints on the Sun and Moon
imposed, dark-time and bright-time observations were then planned
independently.

\begin{table}
\begin{center}
 \begin{tabular}{l|ccc}
  \hline
   & fiducial & min & max\\
  \hline
  MzLS \& DECaLS-z & 100 & 80 & 250 \\
  DECaLS-g & 70  & 56 & 200 \\
  DECaLS-r & 50 & 40 & 175 \\
  \hline
 \end{tabular}
\end{center}
\caption{Exposure times (sec) for DECaLS and MzLS}
 \label{tab:exptimes}
\end{table}

In addition, at all times, we restricted observations to airmass $\le$
2.4 and to pointings that were separated from the Moon by at least
40$^{\circ}$ -- 50$^{\circ}$, with the exact separation determined by the Moon's
phase. We also avoided observing tiles within 1.2$^{\circ}$ of bright
planets.
We enforced minimum and maximum exposure times (see
Table~\ref{tab:exptimes}) to ensure that we
did not exceed depth when observing conditions were excellent, and to
prevent saturation when the sky was too bright and
curtail long exposures in otherwise poor conditions.

The basic logic we adopted is as follows:

\begin{enumerate}
\item Tag tiles with bad exposures as unobserved.
\item Rank order by R.A. and split unobserved tiles by filter.
\item Remove tiles that are too close to the median position of the
  Moon and planets (Mars--Neptune) over the night.
\item Rank order the list of future observing nights, starting with
  the desired night, by local mean sidereal time (LMST) and then split each night into
  one-minute-spaced intervals in LMST.
\item Split the LMST list into dark and bright time.
\item For bright and dark time respectively, match the rank-ordered RA
  and LMST lists by minimizing the time difference between them.
\item Retain LMSTs that are within 5$^{\circ}$ of each R.A.
\item {\it The annealing process}: randomly swap the LMST of two
  tiles. Accept the new positions if the total airmass is
  reduced. Repeat 400 times.
\item {\em For DECaLS only}: prioritize the tiles for building that
  night's plan. Observations are chosen preferentially at decl.
  near decl.$= 0$, with a penalty of 1/100.0 per deg away from
  the equator. Also prioritize selecting tiles near the last
  observation, with a penalty of 1/10.0 per deg for distances more
  than 2$^{\circ}$ away. Priorities are increased (doubled) for observations
  of tiles that have been previously observed in at least one other
  filter. Increase priority for observations of the same tile. This
  should preferentially schedule pairs of $g+r$ exposures in dark
  time. Priorities set to zero for tiles within 1.20$^{\circ}$ of
  Mars--Neptune.
\item Build the plan for the night. Pass 1 is preferentially selecting
  Pass 1 tiles, then Pass 2, then Pass 3, then a repeat observation of an already-observed tile. Pass 2 is
  preferentially selecting Pass 2 tiles, then Pass 3 etc.
\item Observations begin and end at 12$^{\circ}$ twilight for DECaLS and 10$^{\circ}$ for MzLS. 
\item {\it The untangling process}: reduce slews by splitting tiles into
  blocks (consecutive tiles having slews $> 5^{\circ}$) and then trying
  all permutations of the blocks. After this the tiles are split
  again, using blocks of eight consecutive tiles, and the best permutation
  is chosen.
\item Create a list of reserve tiles for bright and dark time from the
  list of observed and unobserved tiles that are closest to transit
  and sufficiently far from the Moon and planets.
\item Observe tiles at their assigned LMST.
\end{enumerate}

\section{Dynamic Observing \label{dynamic}}

\subsection{General Concepts}

Observing conditions at ground-based observatories change due to
temporal and spatial changes in atmospheric transparency and
stability, thermal imbalances between the telescope, dome and ambient
environment, and the spatial location of celestial objects at the time
during which they are observed.

In an ideal world, observing conditions can be monitored during each
on-sky integration as it is in progress, and the total duration of the
ongoing exposure can be modified in real time to ensure that the image
being taken reaches the appropriate depth. This could be accomplished
using, say, nondestructive reads to monitor the actual image data as
it is being collected, or alternatively using some proxy to estimate
the current conditions in the region (e.g., a guide or photometric
camera co-located with the telescope and pointed at the same spot in
the sky).

The hardware realities of the Mosaic3 and DECam instruments prevented
us from implementing any real-time exposure control. However, we were
able to implement the next best option: to analyze each image as soon
as it was taken, estimate the image quality, transparency, resulting
depth and telescope pointing offset, and then adjust our exposure time as soon as
possible, typically with a lag of 1 or 2 images.

At both the Mayall and Blanco telescopes, dynamic exposures were
implemented using two (Python) software ``bots''; both monitored the
observing conditions and telescope pointing offsets, with one ({\tt
  copilot}) providing a graphical view of the derived estimates and
the other ({\tt decbot/mosbot} in the cases of DECam/Mosaic3,
respectively) writing the required scripts and interfacing with the
instrument to modify the exposure time. These codes are all publicly
available\footnote{{\tt https://github.com/legacysurvey/obsbot}}.  We
describe the individual pieces of this process below.

\subsection{Copilot: A Graphical Display \label{sec:copilot}}

\begin{figure}
\begin{center}
 \includegraphics[width=10cm]{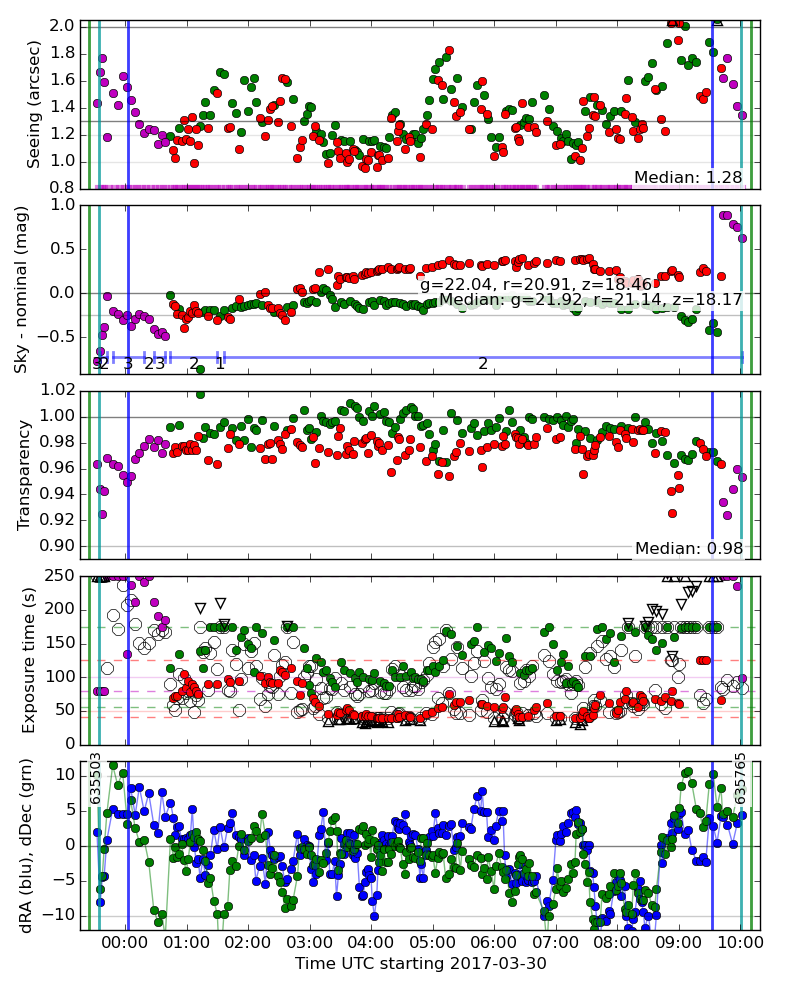}
 \caption{{\tt Copilot} summary plot for DECaLS night of 2017 March 30.}
 \label{fig:copilot}
 \end{center}
\end{figure}

For each raw image, {\tt copilot} measures the seeing, sky brightness,
atmospheric transparency, and photometric zero-point. The bot
extrapolates from the central $1000\times 1000$ pixels of a single CCD or
amplifier (CCD N4 for
DECam and amplifier IM4 for Mosaic3) to infer statistics for the
entire exposure.  For the observers,
{\tt copilot} displays plots of seeing, sky brightness, transparency, and
R.A. and decl. offsets. Figure~\ref{fig:copilot} shows the summary plot
from 2017 March 30.

The combination of {\tt copilot} and either {\tt mosbot} or {\tt
  decbot} performs on-the-fly image reductions, which we describe
briefly below.

\paragraph{Detrending.} The first step is to apply bias and gain,
and then to estimate the sky level by sigma-clipping the central
pixels. This provides a measure of the sky brightness
$m_{\text{sky}}$, assuming the canonical zero-point for the given
camera and filter.

\paragraph{Source detection.} We correlate the image with a matched
filter consisting of a 2D Gaussian with a FWHM of 5 pixels, and
flagging pixels with S/N $\ge 20 \, \sigmasky$. Aperture photometry is
carried out for these (unresolved or star-like) sources using an
aperture with a diameter of 7$\arcsec$ (constant pixel scale) and a sky
annulus with a diameter of 14--20$\arcsec$ (constant pixel scale). The
source counts ($\Ne$) are then counts in the object aperture minus the
mode of sky annulus times the area of the object
aperture.  The following restrictions were applied to ensure a clean sample of sources:
\begin{itemize}
\item $\Ne > 0$;
\item $12 < \mAB <22$;
\item at least 11\arcsec\ (40 pixels) from CCD edges and any other sources;
\item and no bad pixels within 5 pixels of the centroid.
\end{itemize}

\paragraph{Seeing quality determination.} We estimate the seeing by
fitting a circular 2D Gaussian to all sources with $20 < \text{S/N}
<100$, where noise includes the Poisson noise from both the sky and
the source, and only the FWHM is allowed to vary. The seeing we record
is the median of the best-fit FWHM values.

\paragraph{On-the-fly photometric calibration.}
We compute photometric zero-points relative to the PS1 catalogs, and
astrometric offsets from the Gaia DR1 catalogs \citep{gaia,
  gaiaDR1}. Note that we actually use a single PS1--Gaia catalog,
created using a 3.5$\arcsec$ matching radius. There are occasional
holes in the Gaia DR1 catalog in regions that contain plenty of bonafide
PS1 stars, so our astrometry reverts to only using PS1 in such
regions. We enforce the following constraints on the PS1--Gaia catalog:
\begin{itemize}
\item there can only be exactly one match between the catalogs;
\item the PS1 catalog must not flag the source, in the $g$, $r$, and
  \zb bands as coming from a bad CCD region, containing bad pixels, or
  having NaN fluxes;
\item and sources must have a star-like color in the range of $0.4 < g -
  r < 2.7$, where $g - r$ denotes the PS1 median point-spread function
  (PSF) magnitude color.
\end{itemize}

The instrumental zero-point is the difference between the PS1 magnitude
of a source ($\PSmag$) and our measured aperture magnitude ($\mAB$),
and the 2.5$ \sigma$-clipped median for all sources in a CCD,
\begin{align}
\zpt = \text{Med} \lp \PSmag - \mAB \rp + \zpo, \label{eq:zpt}
\end{align}
where $\zpo$ is a band-dependent fiducial zero-point we obtained during
nights with excellent conditions near the start of the DECaLS and MzLS
observations. The relative atmospheric transparency, i.e.\ the
fraction of light that penetrates the Earth's atmosphere relative to a
good night at the start of the survey, can then be computed from the
zero-point,
\begin{align}
\reltransp = 10^{-0.4\lb \zpo - \zpt - \Kco \lp X-1 \rp \rb}~,  \label{eq:transp}
\end{align}
where $\Kco$ is the atmospheric extinction coefficient and $X$ is the airmass.

\paragraph{Depth and exposure factor estimates.}
The 5$\sigma$ AB magnitude depth, with Galactic extinction $\Aco \ebv$ removed, is 
\begin{align}
\mdepth = -2.5 \logten \lp \frac{5 \, \sigmaskyeff}{\texp} \rp + \zpt - \Aco \ebv \label{eq:depth}
\end{align}
where $\sigmaskyeff$ is the square root of sky counts from a region having the size of the source, 
\begin{align}
\sigmaskyeff = \sqrt{ \sigmasky^2 \Neff }\label{eq:sigmaskyeff} \, ,
\end{align}
where $\Neff$ is the noise equivalent area, i.e.\ the effective number
of pixels of an astrophysical source on the CCD, given by
\begin{align}
\Neff = \lp \sum_i v_i \rp^2 / \sum_i v_i^2 \label{eq:Neff} \, ,
\end{align}
where $v_i$ is the
pixelized PSF centered on the source. If the source
is an extended object, then $v_i$ is the value of the PSF convolved
with the object's surface brightness profile.  For speed of computation, the {\tt
  copilot} uses an approximation for $\Neff$ instead:
\begin{align}
\Neffhat &\approx 4\pi \seeing^2 + 8.91 \rhalf^2 + \Ps^2/12 \, , \label{eq:Neff-obsbot} 
\end{align}
where $\rhalf = 0.4\arcsec$ for extended sources and $\rhalf =
0\arcsec$ for point-sources, and $\Ps$ is the
  plate scale. This approximation is based on the assumption that the
seeing is Gaussian, which results in slightly underpredicting the true
value of $\Neff$\ since the seeing profile has larger wings.  In fact, this
approximation  systematically underestimates the true $\Neff$ by
20--40\%, but we have found that a linear model ($A \Neffhat + B$) for
each camera and point-source / extended source pair agrees well with the true
$\Neff$.

Combining equations (\ref{eq:transp})--
   (\ref{eq:sigmaskyeff}), we obtain the exposure time
scaling factor relative to its value under nominal conditions,
 \textit{i.e.} $\ebv=0$,
  $T_{\text{rel}}=1$, and $X=1$,
\begin{align}
\frac{\texp}{\texpo} = \frac{\Neff}{\Neffo} \frac{1}{T_{\text{rel}}^2}
  \times 10^{0.8\lb \Kco \lp X - 1 \rp + \Aco \ebv \rb - 0.4 \lb \msky
  - \mskyo \rb} \, , \label{eq:texpfactor}
\end{align}
where where $\Neffo$ is obtained by using equation (\ref{eq:Neff-obsbot})
with the nominal seeing $\seeing=1.3\arcsec$.
This estimate for the corrected exposure time (rounded up to the
nearest integer) is used to set the duration of the upcoming
exposure.
Finally, the {\tt copilot} compares the depth,
  estimated from detected PS1 stars, attained by a given
image to the desired depth, which is defined as detecting the
canonical 0.45\arcsec\ exponential disk galaxy at S/N=5,
by computing the median $\mdepth$ of all detected sources. The success
factor of the observation is presented as the exposure factor,
$R_{\rm expfac} \equiv t_{\rm observed}/t_{\rm desired}$, i.e., the
ratio between the actual exposure time used for the image and the
exposure time that would have been needed to reach
depth.  The Exposure
Factor is reported on the graph that is visible to the observer.

\begin{figure}
 \includegraphics[width=\textwidth]{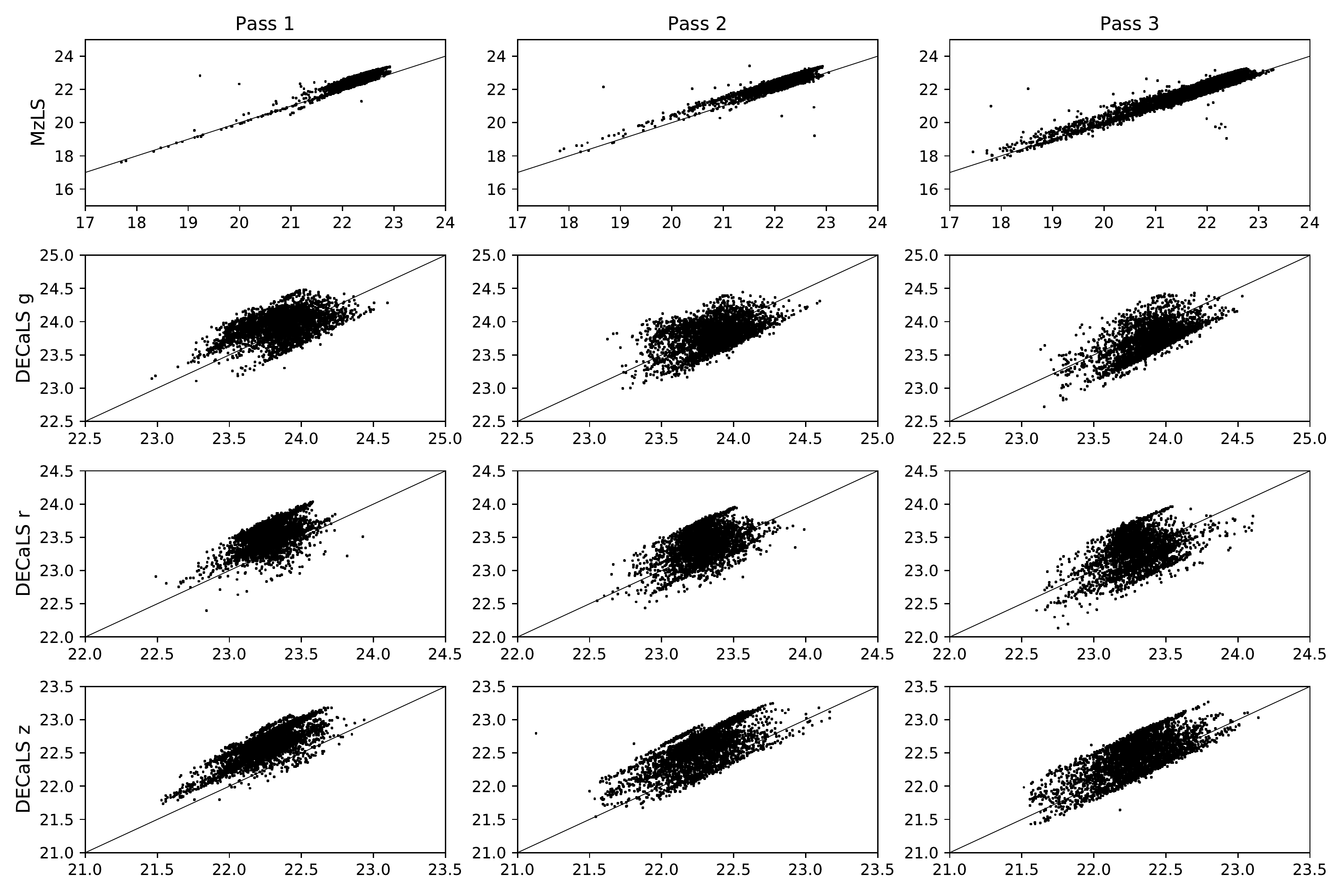}
 \caption{Depth (mag) comparison between the offline pipeline ($x$-axes) and
   {\tt obsbot} estimate ($y$-axes).}
 \label{fig:pipecomp}
\end{figure}

Figure~\ref{fig:pipecomp} shows the comparison between the {\tt
  obsbot} depth estimate and the one determined by the offline
pipeline.  The two are well correlated but is tighter for Mosaic than
DECam.  {\tt obsbot} tends to slightly overestimate the depth, especially for
DECam, but this is a small effect and had no adverse effect for the
overall depth of the surveys.

\subsection{Implementation for MzLS}

For each on-sky exposure that is written to disk, {\tt mosbot}
analyzes a single CCD amplifier to (a) determine the sky level, (b)
detect sources and measure the FWHM, (c) match them to sources from
the PanSTARRS1 catalog, (d) derive the zero-point of the image, (e)
compare this zero-point to the fiducial zero-point to determine the
transparency, and (f) derive the attained depth of the image. In
addition, {\tt mosbot} determines the airmass and Galactic extinction
of the next pointing, predicts the band-dependent seeing based on an
empirical relationship, and calculates
the needed exposure time to reach depth using
Eqn.$\,$\ref{eq:texpfactor}.

{\tt mosbot} only corrects the exposure time for upcoming
observations; the pointing offset of the telescope (which is computed
and displayed by both {\tt mosbot} and {\tt copilot}) has to be
corrected by the night-time observer, and is done while the
exposure is reading out. 

\subsection{Implementation for DECaLS}

At the start of the DECaLS survey, nightly observations began with nominal
exposure times that the observers modified on hour time scales as
conditions changed. On 2015 February 25, we started using {\tt copilot}
and {\tt decbot}. Similarly to {\tt mosbot} for MzLS, {\tt decbot}
uses the most recent raw image on the disk to predict the exposure time
needed to reach the depth at the next pointing.  Tiles with the earliest
LMST are added to the queue while all tiles with LMST in the past are
ignored.

While {\tt decbot} and {\tt mosbot} can also choose the pass number
based on the derived conditions, the observers could force a pass in
conditions that were at the limit between passes. This could be used
to avoid large slews as the surveys progressed, which might have
resulted in larger overheads and uneven completion rates in different passes.

With DECam, the slewing to the next pointing is simultaneous with
reading out the CCD. On average, slewing is faster (about 30~s for
less than 5$^{\circ}$) than read out, so the next exposure usually begins
before the image is built, compressed, and written to the disk. Only after
the image is written can {\tt obsbot} analyze the image and {\tt copilot} provide
an update to the exposure time.  The observing software
cannot change the exposure time once the exposure begins, so the exposure
time would only be updated for the subsequent exposure.

As in the case of MzLS, {\tt decbot} only corrects the exposure time
for upcoming observations; the pointing offset of the telescope has to
be corrected by the night-time observer by temporarily pausing the
exposure queue.

\subsection{Survey Efficiency Gains with Dynamic Observing}
\label{sec:sineff}

\begin{figure}
 \includegraphics[width=9cm]{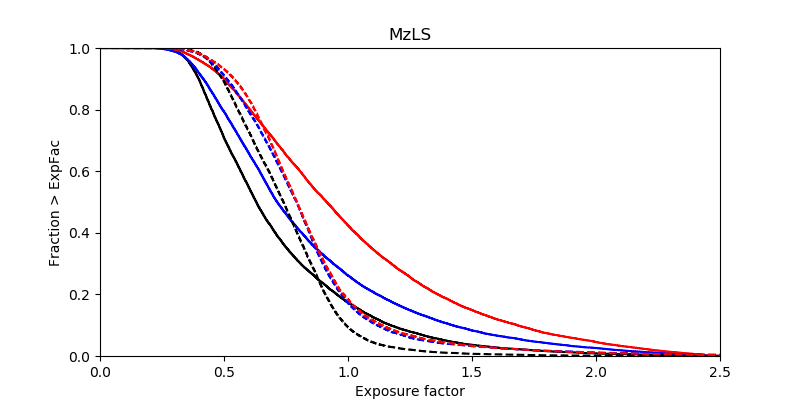}
 \includegraphics[width=9cm]{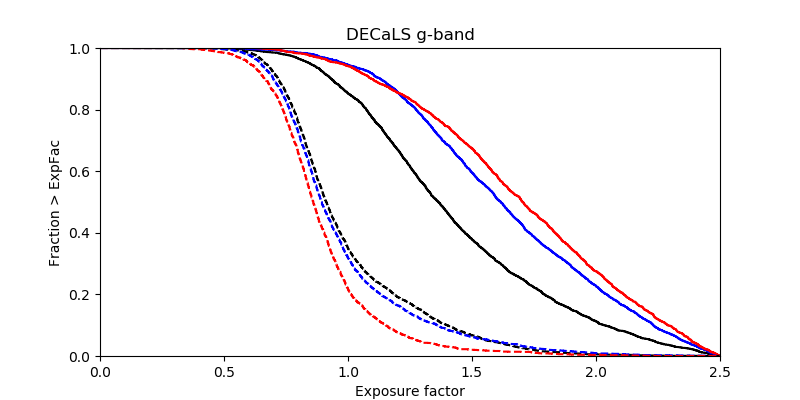}
 
 \includegraphics[width=9cm]{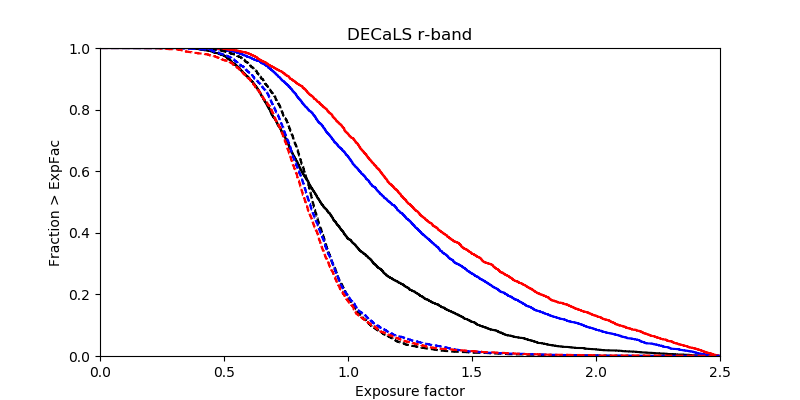}
 \includegraphics[width=9cm]{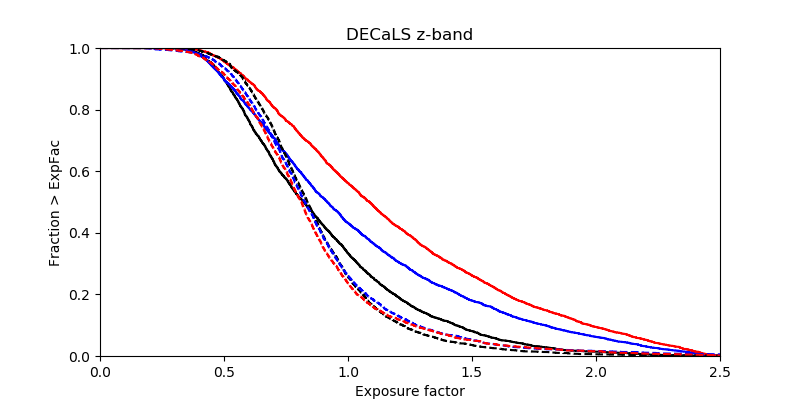}
 \caption{Reverse cumulative distribution of exposure factors for dynamically chosen
   exposure times (solid) and fixed fiducial exposure times under same
   observing conditions (dashed).  Passes 1, 2, and 3 are in black,
   blue, and red, respectively.}
 \label{fig:expfac}
\end{figure}

\begin{figure}
 \includegraphics[width=8cm]{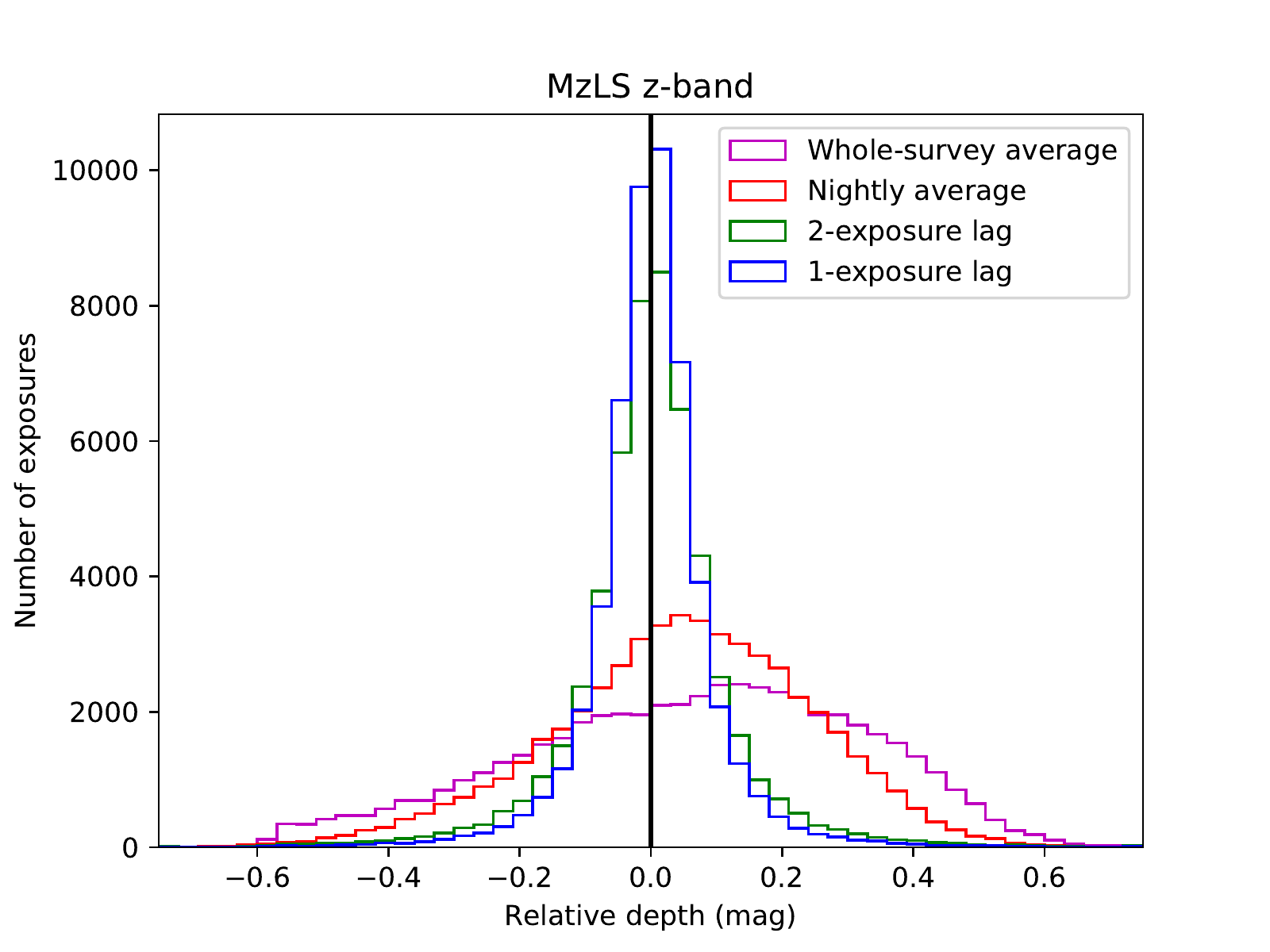}\includegraphics[width=8cm]{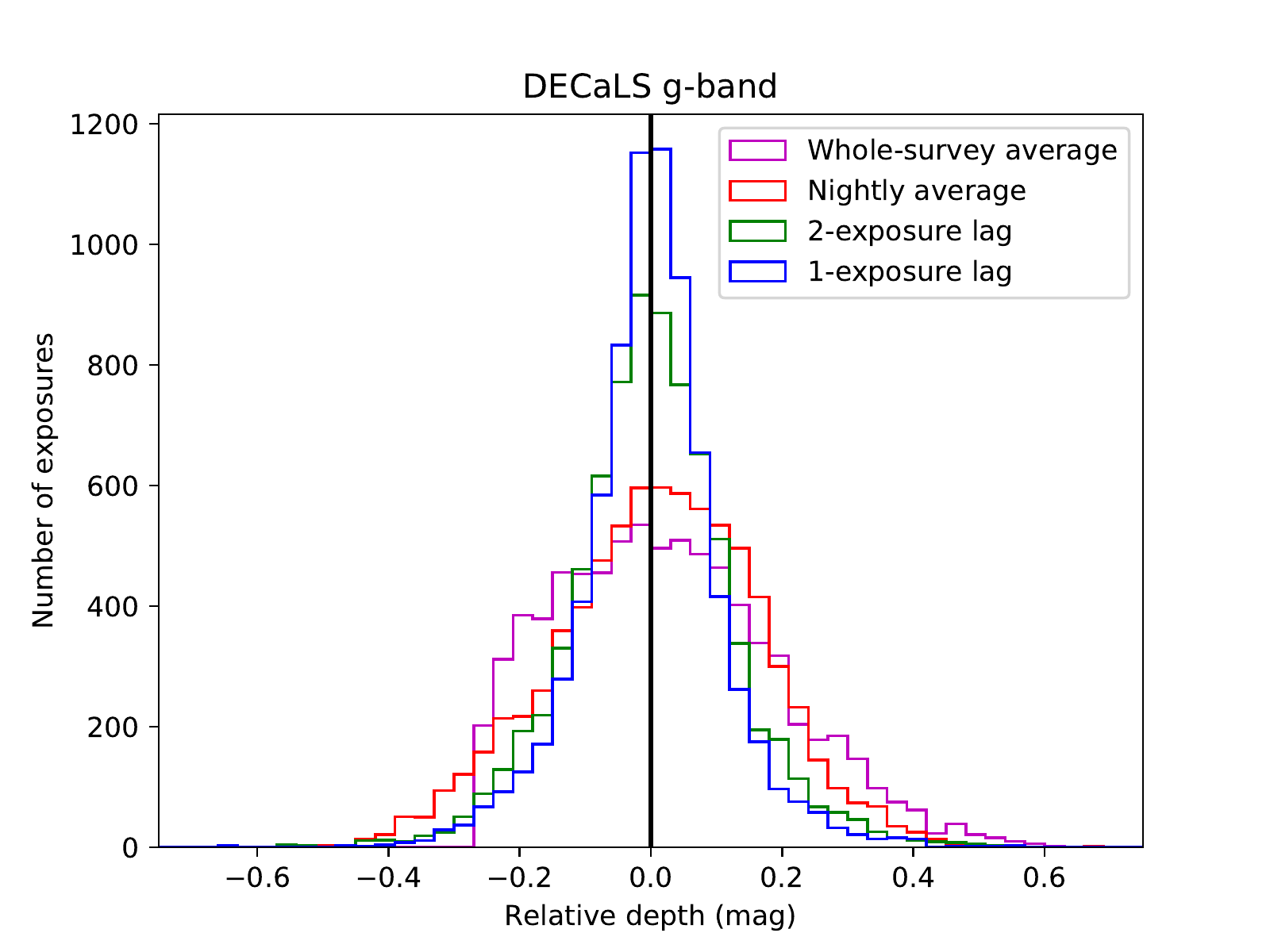}
 
 \includegraphics[width=8cm]{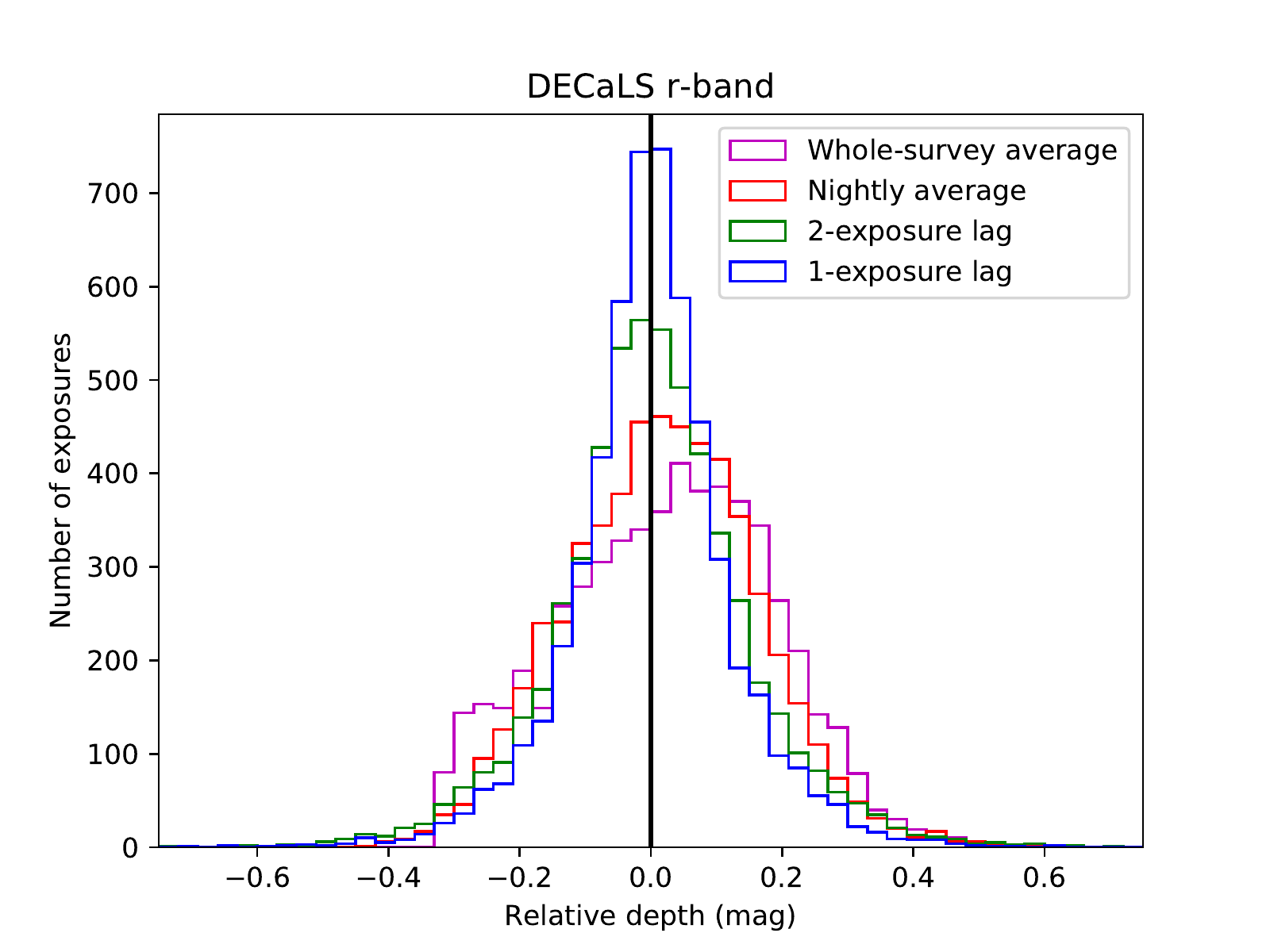}\includegraphics[width=8cm]{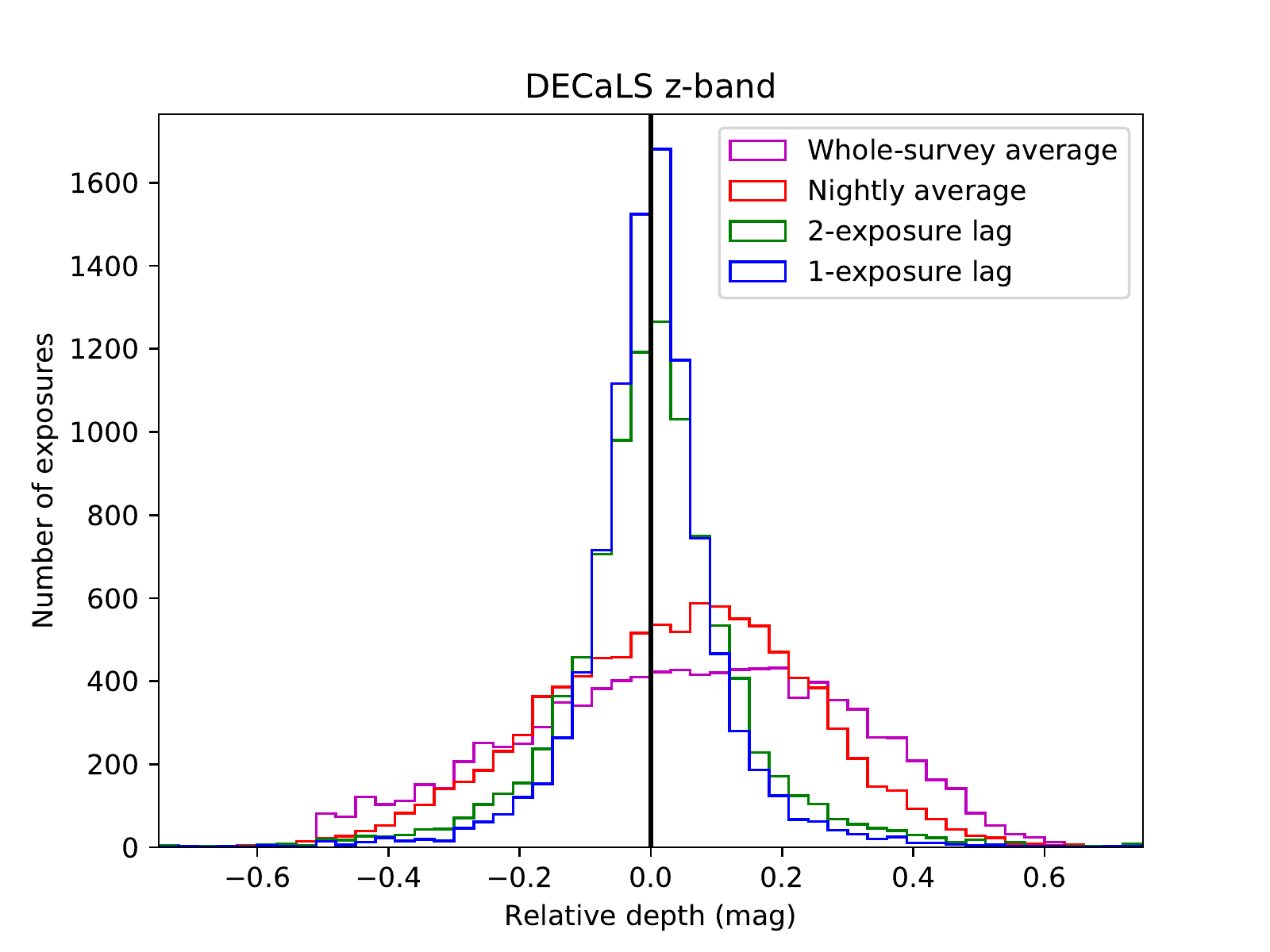}
 \caption{Histograms of relative depth for all dynamically observed
   MzLS and DECaLS images with exposure times within the allowed
   range. Both surveys had a 1--2 exposure lag, so blue and green lines
   show the relative depths we achieved with dynamic exposures. The
   purple and red lines show the relative depth distributions that
   would have resulted had we used a fixed exposure time equal to the
   average needed exposure time for the whose survey (purple) or that
   needed on a nightly basis (red). These latter distributions are
   significantly broader than the green histogram with many more
   outliers demonstrating that a fixed exposure approach would have
   resulted in a much less uniform survey.}
 \label{fig:money-mag-mzls-decals}
\end{figure}

Dynamic exposure times allow the observations to compensate, ideally
in real time, for the variable conditions to ensure that each image
reaches depth. This is demonstrated in Figure~\ref{fig:expfac}, which
shows the reverse cumulative distributions of the exposure factors for the
Pass 1, 2, and 3 images in two cases: the actual MzLS and DECaLS images obtained
under the dynamic exposure time operations, and what would have
resulted if we had used our
fiducial exposure time (see Table~\ref{tab:exptimes}). In the case of the actual observations, we
have restricted our selection to frames with exposure times between
the minimum and maximum times allowed.  In the case of the MzLS
observations, the exposure time was corrected with a typical lag time
corresponding to one frame; for DECaLS, this was two frames, due to
the structure of the queuing software.  Even so, the dynamic observation results
in dramatic gains, especially in the cases of the Pass 2 and 3
observations which are obtained under nonphotometric and/or poor
seeing conditions. In the case of the fixed exposure times, we would
have had to reobserve a larger number of the shallow fields, resulting
in extra on-sky observing time and extra overheads (primarily due to
telescope slews, dome rotations, and CCD readouts).  In addition to
saving time by not underexposing, dynamic observing can save time by
not exposing for longer than is necessary.  This can be seen from
Figure~\ref{fig:money-mag-mzls-decals} which shows the relative depths
of MzLS and DECaLS
exposures using our dynamic exposure strategy versus what we would
have achieved with fixed exposure times, either averaged over the
whole survey or adjusted every night.  The distributions for lags of 1--2
exposures is much narrower around the prescribed depth
than for the two fixed exposure time scenarios, especially in the case
of the $z$ band, for which there are long tails at high relative
magnitudes.

% advantage of dynamic observing 1

\section{Conclusions}
\label{sec:conclusions}

We have presented the overall observing strategy that was used by the
DECaLS and MzLS surveys, for which we implemented a
novel approach of using a dynamic observing strategy, where the
exposure times automatically varied in response to observing conditions
in order to preserve uniformity of survey depth.  We also
implemented a strategy by which every position within the footprints
of these surveys was targeted at least once under photometric
conditions and at least once under conditions of good seeing. This
method results in a demonstrably more uniform survey, which can be
conducted optimally given a finite total amount of observing time and which is
better suited to cosmological studies near the depth limit of a
survey. DECaLS and MzLS are the first surveys to use automated dynamic
exposure times.

Dynamic exposure times may be crucial to future ground-based surveys,
because they conserve telescope time and
increase depth uniformity. The latter also improves searches for transients,
such as moving objects, because non-varying transients
should have a similar probability of detection in images taken at different epochs.

Our method currently necessitates lag times of 1--2
exposures, and removing this constraint would further improve
uniformity of depth.  Amongst surveys currently underway, the
Hobby-Eberly Telescope Dark Energy Experiment (HETDEX) uses
exposure times based on the conditions immediately before the start of
the exposure
(one of us (M.L.), developed an
exposure time calculator and a next field selector for this survey and
we confirmed, through private communication with an active member of
the collaboration, that these were being used during operations).
Based on the success described here, DESI is now planning to implement
real-time dynamic exposures using contemporaneous measures of
the seeing, transparency, and sky brightness.

\acknowledgments

The Legacy Surveys consist of three individual and complementary
projects: the Dark Energy Camera Legacy Survey (DECaLS; NOAO Proposal
ID no. 2014B-0404; PIs: D. Schlegel and A. Dey), the
Beijing-Arizona Sky Survey (BASS; NOAO Proposal ID no. 2015A-0801; PIs:
Z. Xu and X. Fan), and the Mayall \zb-band  Legacy Survey
(MzLS; NOAO Proposal ID no. 2016A-0453; PI: A.D.). DECaLS, BASS,
and MzLS together include data obtained, respectively, at the Blanco
telescope, Cerro Tololo Inter-American Observatory, NSF's
Optical--Infrared Astronomy Research Laboratory (NSF's OIR Lab), the
Bok telescope, Steward Observatory, University of Arizona, and the
Mayall telescope, Kitt Peak National Observatory, NSF's OIR Lab. The
Legacy Surveys project is honored to be permitted to conduct
astronomical research on Iolkam Du'ag (Kitt Peak), a mountain with
particular significance to the Tohono O'odham Nation.

NSF's OIR Lab is operated by the Association of Universities for
Research in Astronomy (AURA) under a cooperative agreement with the
National Science Foundation.

This project used data obtained with the Dark Energy Camera (DECam),
which was constructed by the Dark Energy Survey (DES)
collaboration. Funding for the DES Projects has been provided by the
U.S. Department of Energy, the U.S. National Science Foundation, the
Ministry of Science and Education of Spain, the Science and Technology
Facilities Council of the United Kingdom, the Higher Education Funding
Council for England, the National Center for Supercomputing
Applications at the University of Illinois at Urbana-Champaign, the
Kavli Institute of Cosmological Physics at the University of Chicago,
Center for Cosmology and Astro-Particle Physics at the Ohio State
University, the Mitchell Institute for Fundamental Physics and
Astronomy at Texas A\&M University, Financiadora de Estudos e
Projetos, Fundacao Carlos Chagas Filho de Amparo, Financiadora de
Estudos e Projetos, Fundacao Carlos Chagas Filho de Amparo a Pesquisa
do Estado do Rio de Janeiro, Conselho Nacional de Desenvolvimento
Cientifico e Tecnologico and the Ministerio da Ciencia, Tecnologia e
Inovacao, the Deutsche Forschungsgemeinschaft and the Collaborating
Institutions in the Dark Energy Survey. The Collaborating Institutions
are Argonne National Laboratory, the University of California at Santa
Cruz, the University of Cambridge, Centro de Investigaciones
Energeticas, Medioambientales y Tecnologicas-Madrid, the University of
Chicago, University College London, the DES-Brazil Consortium, the
University of Edinburgh, the Eidgenossische Technische Hochschule
(ETH) Zurich, Fermi National Accelerator Laboratory, the University of
Illinois at Urbana-Champaign, the Institut de Ciencies de l'Espai
(IEEC/CSIC), the Institut de Fisica d'Altes Energies, Lawrence
Berkeley National Laboratory, the Ludwig-Maximilians Universitat
Munchen and the associated Excellence Cluster Universe, the University
of Michigan, the NSF's Optical--Infrared Astronomy Research
Laboratory, the University of Nottingham, the Ohio State University,
the University of Pennsylvania, the University of Portsmouth, SLAC
National Accelerator Laboratory, Stanford University, the University
of Sussex, and Texas A\&M University.

BASS is a key project of the Telescope Access Program (TAP), which has
been funded by the National Astronomical Observatories of China, the
Chinese Academy of Sciences (the Strategic Priority Research Program
"The Emergence of Cosmological Structures" grant no. XDB09000000), and
the Special Fund for Astronomy from the Ministry of Finance. The BASS
is also supported by the External Cooperation Program of Chinese
Academy of Sciences (grant no. 114A11KYSB20160057), and Chinese
National Natural Science Foundation (grant no. 11433005).

The Legacy Survey team makes use of data products from the Near-Earth
Object Wide-field Infrared Survey Explorer (NEOWISE), which is a
project of the Jet Propulsion Laboratory/California Institute of
Technology. NEOWISE is funded by the National Aeronautics and Space
Administration.

The work of E.F.S. was performed under the auspices of the
U.S. Department of Energy by Lawrence Livermore National Laboratory
under Contract DE-AC52-07NA27344.

This research used resources of the National Energy Research
Scientific Computing Center, a Department of Energy Office of Science User Facility
supported by the Office of Science of the U.S. Department of Energy
under Contract No.\ DE-AC02-05CH11231.

The Legacy Surveys imaging of the DESI footprint is supported by the
Director, Office of Science, Office of High Energy Physics of the
U.S. Department of Energy under Contract No.\ DE-AC02-05CH1123, by the
National Energy Research Scientific Computing Center, a Department of Energy Office of
Science User Facility under the same contract, and by the
U.S. National Science Foundation, Division of Astronomical Sciences
under Contract No.\ AST-0950945 to NSF's OIR Lab.

A.D.M. was supported by the U.S. Department of Energy, Office of Science,
Office of High Energy Physics, under Award Number DE-SC0019022.

J.M. gratefully acknowledges support from the U.S. Department of Energy,
Office of Science, Office of High Energy Physics under Award Number
DE-SC002008, and from the National Science Foundation under grant
AST-1616414.

Finally, we would like to thank the anonymous referee, whose comments
helped improve this paper.

\bibliographystyle{aasjournal}  %{mnras} if file is mnras.bst
\bibliography{bib}

%\begin{thebibliography}{}
%
%\bibitem[Astropy Collaboration et al.(2013)]{2013A&A...558A..33A} Astropy Collaboration, Robitaille, T.~P., Tollerud, E.~J., et al.\ 2013, \aap, 558, A33 
%\bibitem[Bertin \& Arnouts(1996)]{1996A&AS..117..393B} Bertin, E., \& Arnouts, S.\ 1996, \aaps, 117, 393 
%\bibitem[Corrales(2015)]{2015ApJ...805...23C} Corrales, L.\ 2015, \apj, 805, 23
%\bibitem[Ferland et al.(2013)]{2013RMxAA..49..137F} Ferland, G.~J., Porter, R.~L., van Hoof, P.~A.~M., et al.\ 2013, \rmxaa, 49, 137
%\bibitem[Hanisch \& Biemesderfer(1989)]{1989BAAS...21..780H} Hanisch, R.~J., \& Biemesderfer, C.~D.\ 1989, \baas, 21, 780 
%\bibitem[Lamport(1994)]{lamport94} Lamport, L. 1994, LaTeX: A Document Preparation System, 2nd Edition (Boston, Addison-Wesley Professional)
%\bibitem[Schwarz et al.(2011)]{2011ApJS..197...31S} Schwarz, G.~J., Ness, J.-U., Osborne, J.~P., et al.\ 2011, \apjs, 197, 31  
%\bibitem[Vogt et al.(2014)]{2014ApJ...793..127V} Vogt, F.~P.~A., Dopita, M.~A., Kewley, L.~J., et al.\ 2014, \apj, 793, 127  
%
%\end{thebibliography}



%\facilities{KPNO:Mayall (Mosaic3), Steward:Bok (90Prime), CTIO:Blanco (DECam), WISE, NEOWISE, Gaia}
\facility{KPNO:Mayall (Mosaic3), Steward:Bok (90Prime), CTIO:Blanco
  (DECam), WISE, Gaia.}

\end{document}